\documentclass[twocolumn,trackchanges,twocolappendix]{aastex63}
\usepackage{lineno}
\usepackage{amsmath}
\usepackage{graphicx}
\usepackage{epstopdf}
\usepackage{color}
\definecolor{marine}{RGB}{255,0,127}

\received{February 29, 2021}
\revised{February 30, 2022}
\accepted{February 31, 2023}
\submitjournal{ApJ}

\shorttitle{A Red Giants' Toy Story}
\shortauthors{Miller Bertolami, M. M.}

\begin{document}

\title{A Red Giants' Toy Story}

\correspondingauthor{Marcelo M. Miller Bertolami}
\email{mmiller@fcaglp.unlp.edu.ar, marcelo@mpa-garching.mpg.de}

\author[0000-0001-8031-1957]{Marcelo M. Miller Bertolami}
\affiliation{Instituto de Astrof\'isica de La Plata, Consejo Nacional de Investigaciones Cient\'ificas y T\'ecnicas \\
Avenida Centenario (Paseo del Bosque) S/N,  B1900FWA La Plata, Argentina.}

\affiliation{Facultad de Ciencias Astron\'omicas y Geof\'isicas,  Universidad Nacional de La Plata\\
  Avenida Centenario (Paseo del Bosque) S/N,  B1900FWA La Plata, Argentina.}



\begin{abstract}
In spite of the spectacular progress accomplished by stellar
  evolution theory, some simple questions remain unanswered. One of
  these questions is ``Why do stars become red giants?''. Here we
  present a relatively simple analytical answer to this question. We
  validate our analysis by constructing a quantitative toy model of a
  red giant and comparing its predictions to full stellar evolutionary
  models.

We find that the envelope forces the value of $\nabla=d \ln T/d
  \ln P$ at, and above, the burning shell into a very narrow range of
  possible values. Together with the fact that the stellar material at
  the burning shell both provides and transports most of the stellar
  luminosity, this leads to tight relations between the thermodynamic
  variables at the burning shell and the mass and radius of the core---$T_s(M_c,R_s)$, $P_s(M_c,R_s)$, and $\rho_s(M_c,R_s)$.
  When complemented by typical mass-radius relations of the helium
  cores, this implies that for all stellar masses the evolution of the
  core dictates the values of $T_s$, $P_s$ and $\rho_s$. We show that for
  all stellar masses evolution leads to an increase in the pressure
  and density contrasts between the shell and the core, forcing a huge
  expansion of the layers on top of the burning shell.

Besides explaining why stars become red giants our analysis also offers a  mathematical demonstration of the
 so-called shell homology relations, and provides simple quantitative answers to some properties of low-mass red giants.

\end{abstract}



\section{Introduction} \label{sec:intro}

About a century from its humble beginnings
\citep{1926ics..book.....E}, stellar evolution theory has developed
into a full-fledged predictive theory \citep{2012sse..book.....K}
whose results are routinely used as inputs for other fields of
astrophysics \citep{2011spug.book.....G,2013osp..book.....C}. Its
predictions have been confirmed by a variety of different
observational tests, and are continuously checked by countless
numerical simulations by means of specifically tailored numerical
codes \citep{2008Ap&SS.316....1L,2020A&A...635A.164S}.  In spite of
the spectacular progress accomplished, some simple questions remain
unanswered. One of these questions is ``Why do stars become red
giants?''. Although the existence of a red giant solution is known
since the early days of automatic numerical computations
\citep{1938PTarO..30D...1O,1947PThPh...2..127H,1949PhRv...75.1619H,1955ApJS....2....1H},
no simple explanation of the reason for the existence of such solution
has been provided. The absence of a universally accepted answer for
this question can be clearly appreciated by reading the related
chapters in textbooks of stellar evolution. When trying to explain the
evolution after the end of core-hydrogen (H) burning, different authors
choose very different paths. Some authors choose to describe
extensively the details of numerical models
\citep{2013sepa.book.....I}, others choose to establish an ad hoc
principle that applies exclusively to stellar evolution
\citep[e.g. the so-called ``mirror principle of radial
  motions''][]{2012sse..book.....K}, others try to convince themselves
that not all physical processes can be understood in simple
step-by-step terms and that we should accept the raw output of
numerical simulations \citep{2009itss.book.....P}, while others openly
admit that we lack a definitive explanation of the precise physical
reason(s) that drive the expansion
\citep{2004sipp.book.....H,2005essp.book.....S}.

The lack of an accepted answer does not mean that no answers have been
proposed but just that proposed answers have never attained wide
popularity. And for good reasons. Proposed answers encompass a wide
  variety of ideas, too vast and complex for this introduction. With
  no intention of being exhaustive, but to highlight both the
  complexity of the problem and its many possible angles, let us
  mention a few of them. As mentioned by \cite{1981ASSL...88..179E}, one
  of the most ubiquitous myths among nonspecialists is the idea that
  the gravitational energy released by the contracting core is
  absorbed by the envelope causing it to expand, ignoring the fact
  that most of the energy comes from the burning shell and the
  envelope has no way to tell were the energy is coming
  from. Similarly ubiquitous is the statement that the increase in the
  stellar radius is a consequence of the increase in the energy output of the
  burning shell due to the heating of the core, also ignoring the fact
  that in low-mass stars the core is mostly heated by the surrounding
  envelope (and not the other way around), and that there is no
  obvious reason why the burning shell would not settle at a 
  lower temperature, which would imply a huge decrease in the energy
  output of the burning shell. While it might be expected
  that nonspecialists  find it hard to pin down the reasons why
  stars become red giants, it is much more interesting that experts
  in stellar structure have also failed to agree on the actual causes
  of this transformation.  While studying this problem some authors
  have highlighted the importance of the gravitational field generated
  by the compact core by numerically solving steady-state 
  solutions (thermal equilibrium, $dS/dt=0$)
  \citep{1973A&A....25...99H,1981ASSL...88..179E,1983A&A...127..411W}. Others
  have focused on the chemical gradients developed due to nuclear
  reactions at the burning shell \citep{1949MNRAS.109..614H}. Some
  authors have highlighted the key importance of the degree of central
  condensation of the core and its consequences on the
  envelope \citep{1991ApJ...383..757E,2005slfh.book..149F}. While all
  the works mentioned before have been focused on the study of the
  consequences of different core properties in steady-state solutions,
  some other authors have concluded that giantness is a consequence of
  thermal instabilities in the envelope due to the high luminosity
  released by the burning shell \citep{1984IAUS..105...21R,
    1988ApJ...329..803A, 1992ApJ...400..280R,
    1994ApJ...433..293R}. Also in this connection, some authors have
  pointed to the development of some gravothermal catastrophe of the
  core and the entropy gradient above it
  \citep{1991ApJ...374..631F}. The methods adopted to try to answer
  this question have also been extremely diverse. For example, some
  authors have tried to answer the question by constructing simplified
  ad hoc polytropic models
  \citep{1991ApJ...383..757E,1992PASA...10..125F,
    1998MNRAS.298..831E}, some have tried to answer this question by
  means of dimensional analysis \citep{1991ApJ...372..592B}, while
  others have tried to answer the question by a detailed evaluation of
  numerical solutions on the $U-V$ plane \citep{1991ApJ...374..631F,2000ApJ...538..837S}. While some authors
  have highlighted the importance of some critical curves and singular
  points in the $U-V$ plane \citep{1985ApJ...296..554Y,
    2000ApJ...538..837S}, others have disparaged these explanations as
  mere descriptions of the results from numerical models
  \citep{1993ApJ...415..767I}. Proposed answers also encompass a wide
  diversity in complexity, from very simple
  \citep{2000AmJPh..68..421H} to extremely complex
  \citep{10.1093/mnras/236.3.505}.
The discussions about why stars
become red giants have sometimes turned into heated debates
\citep{1997seas.conf...19S,1997seas.conf....9F,2000ApJ...538..837S,
  2005slfh.book..149F}, while some other times authors have ignored
criticism and continue to develop ideas \citep{1992ApJ...400..280R,
  1994ApJ...433..293R} that had already been seriously questioned by
other researchers \citep{1989A&A...209..135W,1993ApJ...415..767I}.
The reader interested in the subtleties and shortcomings of the different ideas is referred to the appendix of \cite{2000ApJ...538..837S}, Appendix 10.B of \cite{2005slfh.book..149F}, the introductions to \cite{1988ApJ...329..803A} and \cite{2009PASA...26..203S}, and Section 3.3 of \cite{1991ApJ...372..592B}.

Both \cite{10.1093/mnras/236.3.505} and \cite{1993ApJ...415..767I} concluded that a simple explanation of why stars become red giants is not possible.
\cite{1993ApJ...415..767I} issued a warning against attempts to
find simple explanations, claiming that simple explanations might be
forcefully misleading.   Although we are aware of these warnings,
this paper is an attempt to shed some light on a possible simple
explanation on the question of why do stars become red giants. We
understand this to be a worthwhile project because toy models and
dimensional analysis play a key role in how our minds approach the
understanding of physical problems. We believe that toy models are key
to extrapolate the results of a set of numerical simulations to all
\emph{similar} stars and to other \emph{analogous} physical problems. In
fact, in order to define what the words \emph{similar} and \emph{analogous} in the previous sentence mean, a toy model description of
the problem is a necessity.  For this reason we have searched for a
toy model description of red giants that is both accurate and
convincing.  In constructing our toy model we have kept in mind the
warnings by \cite{1993ApJ...415..767I} regarding ``convoluted
arguments'' and ``circular explanations''. We have forced ourselves to
avoid the use of particular properties observed in detailed stellar models
in the construction of our toy model, and kept the line of reasoning
as clear as possible.

In our opinion, numerical experiments with steady state solutions
  (thermal equilibrium) like those performed by
  \cite{1973A&A....25...99H}, \cite{1983A&A...127..411W}, and
  \cite{1993ApJ...415..767I}, together with the fact that low-mass red
  giants evolve in a nuclear timescale and develop the most extreme
  case of giantness, clearly show that thermal instabilities in the
  envelope are not what pushes stars into red giant dimensions. In
  fact, as pointed out by \cite{1997seas.conf....9F}, even when they
  expand in thermal timescales envelopes lag behind, rather than lead,
  as can be tested by shutting off chemical changes in a stellar
  evolution code and letting the envelope reach its final steady-state
  solution consistent with the structure of the core.  One of the
most insightful presentation to date was given by
\cite{2005slfh.book..149F} who describes {\it giantness} as a
structural property of thermal equilibrium solutions
 consisting of a
compact core, the presence of a burning shell, and a massive
  envelope that imposes a "mass storage problem" to the star. The fact
  that it is a structural property of thermal equilibrium solutions
  does not imply that a star out of thermal equilibrium cannot be in a
  giant configuration, but just that in those cases we should
  understand them as evolving toward the corresponding structure in
  thermal equilibrium in a Kelvin-Helmholtz timescale. In his article,
  \cite{2005slfh.book..149F} proceeds first by defining a set of
  principles of stellar structure and the meaning of a dense
  core. Then, assuming a simple model with a discontinuity at the
  burning shell, proceeds to derive an ``asymptotic theory'' of low-mass
  red giants. For this, he first motivates that the temperature at
  the discontinuity follows a very specific relation (his equation
  10.2) with the help from polytropic envelope integrations (performed
  in his Appendix 10.A.2). From these expressions, and using some
  numerical results of polytropic envelope integrations, he argues
  that massive envelopes have forcefully a very small value of his
  parameter $\tau$, which effectively means that temperature at the
  burning shell has a tight dependence on the mass and radius of the
  core. Using this result, he then derives two "theorems'' for how the
  luminosity of the star and the density of the shell relate to the
  mass and radius of the core---his equations 10.7 and 10.8; the
  latter only valid for a very specific temperature dependence of the
  energy generation rate. Once this is done he proceeds to extend his
  results to a unified treatment of main-sequence stars and red
  giants, horizontal branch stars, and finally, to giant stars with
  luminous cores. However, we understand the work by
\cite{2005slfh.book..149F} to be problematic due to a florid but
  mathematically obscure presentation, and the reliance on
assumptions difficult to justify. Among these, his derivation of
  his equations 10.5 and 10.18 stick out, as their derivation require
  that $P_{\rm rad}/P_{\rm gas}$ is constant throughout the burning
  shell, but nonzero, imposing a relation between $T^4$ and $P$ which
  disappears when radiation pressure is negligible\footnote{Which
    implies that $1-\beta$ is close to zero. Oddly enough radiation
    pressure is negligible in the conditions relevant for red giants,
    and the appearance of the $1-\beta$ factor is rather
    disturbing.}. Another feature that makes his equation 10.5
    suspicious is the fact that the energy generation from the burning
    shell is proportional to the cube of the radius of the burning
    shell, while in a thin shell approximation the expected dependence
    would be with the square of the radius. More worrying is the
  fact that, within his asymptotic theory, the core-mass-luminosity relation
  (his equation 10.7) is obtained without any mentioning regarding the
  development of convection in the envelope. This result is a
  consequence of the assumption, in his asymptotic theory, that
  $\tau=0$, which is not rigorously demonstrated and only based on the
  discussion of polytropic envelopes on the $U-V$ plane. In fact, it
  has been known since the work of \cite{1952ApJ...116..463S} that
  without the inclusion of an outer convective region, post main
  sequence models expand without increasing their luminosity (see for
  example Fig. 3 in \citealt{1952ApJ...116..463S} and Fig. 2 of
  \citealt{2009PASA...26..203S}). This issues, together with his
  assumption of a very specific temperature dependence for the
  CNO-Cycle ($\eta\simeq 15$, his equations 10.8 and 10.10), his use
  of the $U-V$ plane (in his Appendix 10.A.2), and his emphasis on
  polytropic envelopes (particularly the $n=3$ polytrope), makes his
  final conclusions difficult to accept.  In spite of these
  mathematical shortcomings, we believe his perceptive physical description
  of the red giant structure is worth exploring.

The aim of this paper is to present a compelling toy model of red
giants that improves our insight of how these stars work. Note in
passing that the key feature of toy models is not precision but the
accurate description of the main processes involved\footnote{``All
  models are wrong but some are useful'' \cite{Box1979}.}, albeit
distilled to its most simple elements. This is the ideal we pursue
through the paper.  The paper is structured as follows; first on
Section \ref{sec:shell} we analyze the burning shell and its
surroundings. This sets the key constraints imposed by the burning
shell and the envelope on the thermodynamic variables. Moreover, this
section offers the first proof we are aware of for the underlying
hypothesis of the so-called shell homology relations; i.e. that
temperature ($T_s$), pressure ($P_s$), density ($\rho_s$) at the
location of the burning shell only depend on the radius ($R_s$) and
mass ($M_c$) of the core. In addition, these sections will clarify the problems with some assumptions behind Faulkner's asymptotic theory and highlight the role of convection in the outer layers. Then on Section \ref{sec:toy-lowmass} we
check the accuracy of the derived expressions by constructing a toy
model of a full low-mass red giant and compare its predictions with
those of detailed stellar evolution computations.  Interestingly this
toy model helps us understand why all low-mass stars develop the
Helium flash at nearly the same core mass, why this mass is about a
half a solar mass and why the burning shell in low mass stars remains
at an almost constant location through the evolution. After this
validation, on Section \ref{sec:dym} we make a dimensional analysis of
red giant stars of all masses. Dimensional analysis of the resulting
expressions shows why the presence of a burning shell makes homologous
contraction of the whole star impossible and forces the
formation of bright red giants. A simple description of the formation
of red giants at all masses is presented on Section
\ref{sec:toystory}. We conclude the paper by summarizing our results
and the physical insights gained from the present model. We hope that
the last two sections will convince our most numerically minded
readers of why simple models can improve our understanding of the
inner workings of stars, even after so many decades of numerical
simulations.


\section{The burning shell and its surroundings} \label{sec:shell}
\begin{figure}
\centering
\includegraphics[width=\columnwidth]{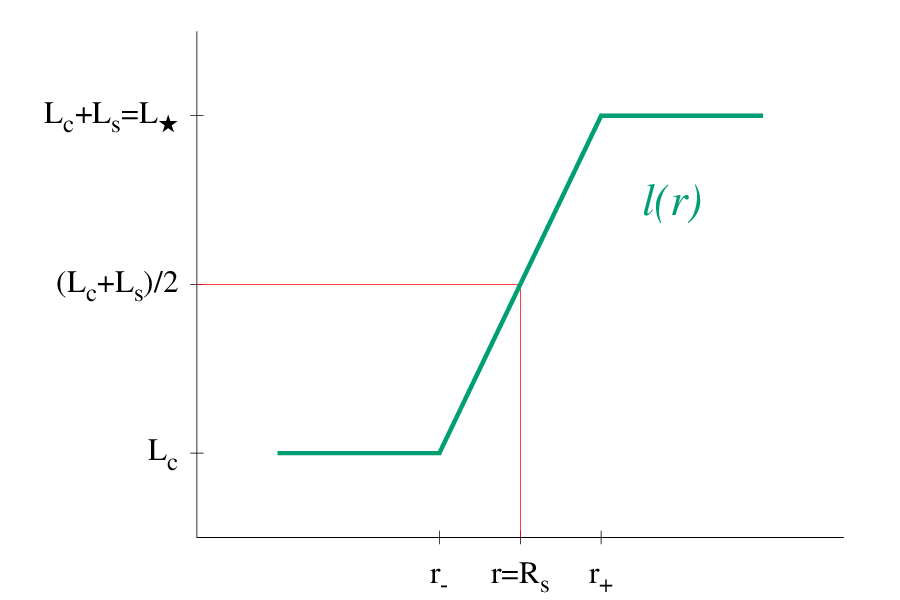}
\caption{Schematic description of the evolution of the luminosity
  $l(r)$ around the burning shell for a core with generic luminosity
  $L_c$. Note that $r_{-}$, and $r_{+}$ indicate the position of the
  inner and upper boundaries of the burning shell, and $L_s$ indicates
  the total energy per unit time generated in the burning shell.}
\label{fig:shell}
\end{figure}

One of the defining features of a red giant star is the presence of a
very localized burning shell.
Fig. \ref{fig:shell} shows a schematic description of the region of
the star around the burning shell\footnote{Note that our treatment of the burning shell is rather different from that of \cite{2005slfh.book..149F}. In that work the burning shell was adopted as a discontinuous transition making no distinction between the values of the variable below, at and above the burning shell. In view that some of these variables can have significant changes (e.g. $L_c\ll L_s$) we think that approximation is not justified.}. We will indicate by $T_-$, $P_-$,
$\rho_-$ and $r_-$ the structure variables at the bottom boundary of the
burning shell, and by $T_+$, $P_+$, $\rho_+$, and $r_+$ those at the
upper boundary of the burning shell. Energy generation happens between
$r_-$ and $r_+$ where the luminosity $l$ increases from the luminosity
of the core ($L_c$) to the luminosity of the envelope
($L_c+L_s$). $R_s$ indicates the location of the middle of the burning
shell, where the thermodynamic variables are $T_s$, $P_s$,
$\rho_s$. Note that for a burning shell to have an impact on the
structure of the star its luminosity must be at least similar to that
of the core ($L_s\gtrsim L_c$). In most interesting cases in fact
$L_s$ will be $L_s\gg L_c$.
Due to the extreme sensitivity of
nuclear reaction rates with the temperature (e.g. $\epsilon_{\rm
  CNO}\simeq \epsilon_0 \rho^t T^\nu$, with $t=1$, and $\nu\simeq
23\textendash 10$ for $T\simeq 10^7 \textendash 10^8$K), a small
decrease in the temperature is enough to produce a huge drop in the energy
generation rate. A small decrease of $\delta T/T \simeq 1/\nu$ leads the
energy release to drop almost to zero $\delta\epsilon/\epsilon\simeq
\nu\, \delta T/T\simeq 1$. As a consequence the energy released by the
burning shell is concentrated in a thin region of $\sim 2 \delta T$ around
the characteristic temperature of the shell $T_s$. This will allow us
to work with first-order approximations around the shell temperature
$f(T_s+dT)\simeq f(T_s) + \frac{df}{dT}(T_s)\, dT$ and consider the burning shell to be thin for most computations (i.e. $r_-\simeq R_s\simeq r_+$).

Due to the presence of the H-burning shell, the stellar plasma at the
shell cannot be degenerate\footnote{Otherwise a flash would develop,
injecting huge amounts of energy in that layer, and lifting
degeneracy.}. Then, as the material on the burning shell is not
degenerate, and $T$, $\rho$, and $P$ are continuous, there {\it must}
exist regions, immediately below and above the burning shell, where the
material behaves as a classical ideal gas ($P=\Re\, \rho\,
T/\mu$)\footnote{Strictly speaking as a classical ideal gas plus
radiation, but radiation pressure always plays a minor role at the
burning shell. In fact, we know from observations that most giant stars
are far from the Eddington luminosities of their cores.}.

\subsection{Above the burning shell} \label{sec:aboveshell}
As a first step on our way to understanding the physics of red giants,
we will look at how the outer boundary conditions establish some
constraints on the thermodynamical quantities immediately above the
burning shell. In particular, we will see that the outer boundary
conditions strongly restrict the possible values of the temperature
gradient $\nabla={\mathrm d} \ln T/{\mathrm d} \ln P$.

Under the assumption that the envelope is in a steady state (thermal
equilibrium, $dS/dt\simeq 0$, $\int_0^{M_\star} |\epsilon_g| dm\ll
L_\star$), the thermodynamic variables in the envelope above the
burning shell must fulfill the hydrostatic equilibrium equation,
\begin{equation}
  \frac{dP}{dm}=-\frac{G m}{4\pi r^4},
  \label{eq:hydro0} 
\end{equation}
the connection between the local radius and the lagrangian mass coordinate,
\begin{equation}
  \frac{dr}{dm}=-\frac{1}{4\pi r^2\rho},
  \label{eq:radio0} 
\end{equation}
and the heat transport equation,
\begin{equation}
  \frac{dT}{dm}=-\frac{G m T}{4\pi r^4 P} \nabla.
  \label{eq:transport0} 
\end{equation}
In the last equation $\nabla$ is given by
\begin{equation}
  \nabla=\nabla_{\rm rad}=\frac{3}{16\pi ac G} \frac{\kappa L_\star P}{ m T^4},
  \label{eq:nablarad} 
\end{equation}
when $\nabla_{\rm rad}<\nabla_{\rm ad}$ and energy is transported by radiation, or $\nabla=\nabla_{\rm ad}$ when convection develops \citep{1906WisGo.195...41S}.
All quantities in eqs. \ref{eq:radio0}, \ref{eq:transport0}, and \ref{eq:nablarad}   have been defined as in the classic textbook by
\cite{2012sse..book.....K}, and $L_\star=L_s+L_c$ is the luminosity of the star.

Given that we want to understand the general properties imposed by the
envelope on the burning shell for a core of arbitrary mass ($M_c$) and
radius ($R_s$) it is convenient to use adimensional variables.
Using the mean pressure of the core
$\bar{P_c}$, the mean density of the core $\bar{\rho_c}$ and a characteristic temperature $T_c$\footnote{Defined as the temperature of a classical ideal gas with the mean hydrostatic pressure $\bar{P_c}$ and mean density $\bar{\rho_c}$ of the core.},
\begin{equation}
  \bar{P_c}=\frac{G {M_c}^2}{8\pi {R_s}^4},\ \ \ \ \ \ 
  \bar{\rho_c}=\frac{3}{4\pi}\frac{M_c}{{R_s}^3},\ \ \ \ \ \
  T_c= \frac{\mu G M_c}{6 \Re R_s}
\label{eq:characteristic_PrhoT}
\end{equation}  
we define $q=m/M_c$, $x=r/R_s$, $y=P/\bar{P_c}$, $z=\rho/\bar{\rho_c}$ and $t=T/T_c$.
Assuming that the gas is a classical ideal gas and that the opacity can be approximated by a power law
$\kappa=\kappa_0 P^a T^b$($a\geq 0$, $b\leq 0$ under normal
 conditions), eqs. \ref{eq:hydro0}, \ref{eq:radio0}, \ref{eq:transport0}, \ref{eq:nablarad}  can be written as
\begin{equation}
\frac{dy}{dq}=\frac{-2q}{x^4},\ \ \ \ \ \ \frac{dx}{dq}=\frac{t}{3x^2y},\ \ \ \ \ \ \frac{dt}{dq}=\mathbb{C} \frac{y^a t^{b-3}}{x^4},
\label{eq:adimensional}
\end{equation}  
and
\begin{equation}
  \nabla=\min\left(\mathbb{C} \frac{y^{a+1} t^{b-4}}{-2q}, 0.4 \right).
  \label{eq:nabla}
\end{equation}
The meaning of the constant $\mathbb{C}$ ($\mathbb{C}<0$) can be made clear by defining a characteristic luminosity\footnote{The reader is warned not to confuse $L_0$, which is a reference luminosity for a core of given mass and radius, with the luminosity of the core $L_c$ to be defined in the next sections.} $L_0$
for a core of mass $M_c$ and radius $R_s$ as
\begin{equation}
  L_0=\frac{64 \pi^2 a c}{3} \frac{{T_c}^4{R_s}^4}{M_c \kappa_c},
  \label{eq:L0}
\end{equation}
where $\kappa_c=\kappa(\bar{P_c}, T_c)$. With this definition
it becomes clear that $\mathbb{C}$ is an adimensional version of the
luminosity of the star and $\mathbb{C}=-L_\star/L_0$.

Due to hydrostatic equilibrium pressure at the burning shell is necessarily lower than the mean pressure at the core, and even more so at the top boundary of the burning shell ($y_+<1$, and much lower than unity for dense cores). On the other hand, as temperature does not change significantly across the burning shell it is usually $t_+\lesssim 1$. Note that as $T_c$ corresponds to the temperature required by a classical ideal gas at the mean density of the core to match the hydrostatic pressure, for degenerate cores $T_c$ is actually larger than the real temperature of the core and thus of the shell, and then $t_+<1$. We can get a clear idea of the relevant values of $\mathbb{C}$ by replacing the quantities in eq. \ref{eq:L0} with those of a typical envelope ($\mu\simeq 1.176$, $\kappa\simeq 0.34$cm$^2$g$^{-1}$) and obtain,
\begin{equation}
  L_0= 6.77\times10^{35} \left(\frac{M_c}{M_\odot}\right)^3 \hbox{erg/s}=177\left(\frac{M_c}{M_\odot}\right)^3 L_{\odot}
  \label{eq:L0_bis}
\end{equation}
We see from eq. \ref{eq:L0_bis} that for typical values of core masses in red giants, $L_0$ is lower than the typical luminosities of the same stars on the main sequence. Physically interesting values of $\mathbb{C}$ correspond to $\mathbb{C}\gtrsim 1$ and even $\mathbb{C}\gg 1$ in the case of low-mass stars.

\begin{figure}
\centering
\includegraphics[width=\columnwidth]{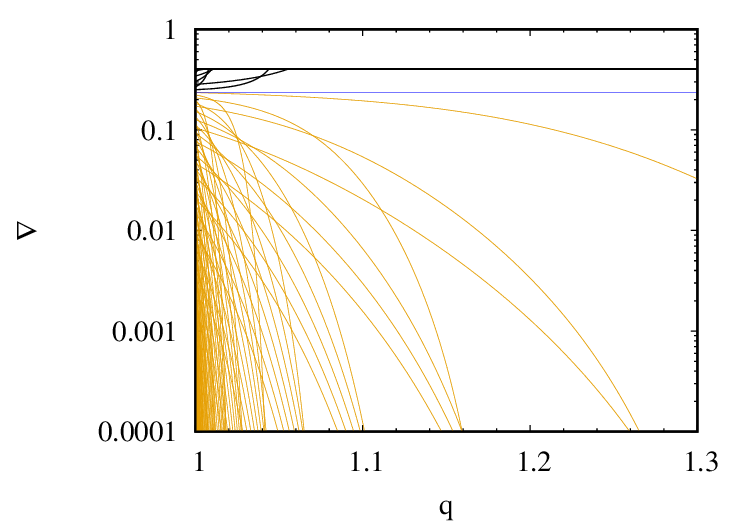}
\includegraphics[width=\columnwidth]{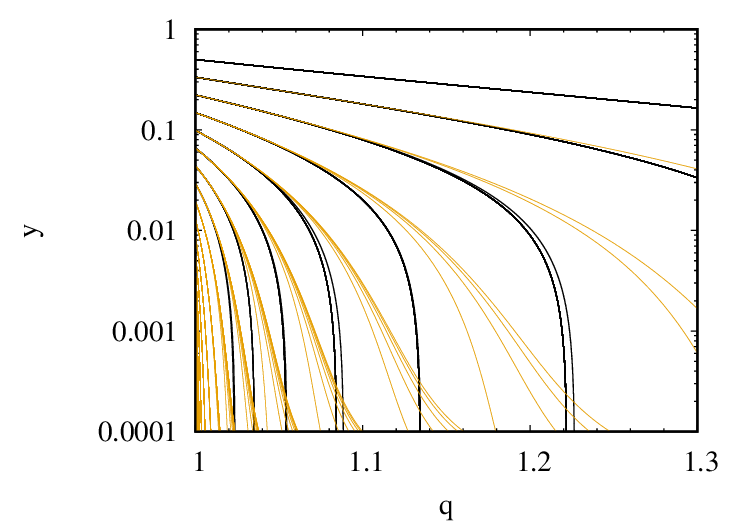}
\includegraphics[width=\columnwidth]{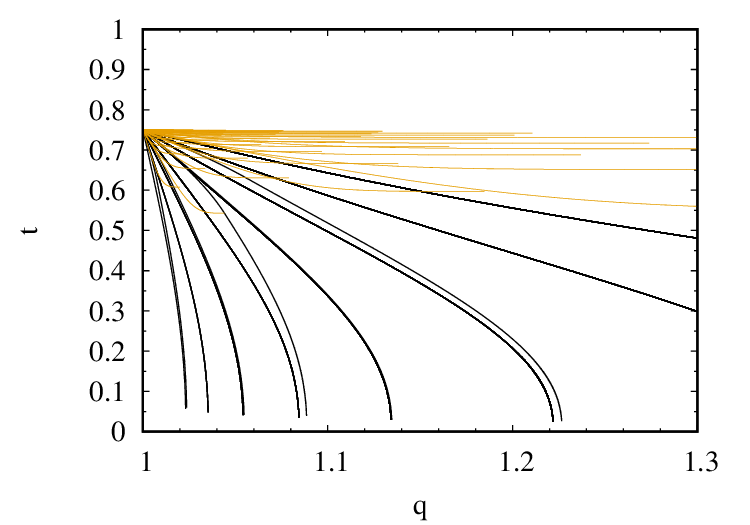}
\caption{Envelope integration for different values of $\mathbb{C}$ and $y_+$, for $t_+=0.75$ and for a Kramers' opacity.}
\label{fig:envolturas}
\end{figure}

Using eqs. \ref{eq:adimensional} we can integrate the envelopes from
the upper boundary of the burning shell ($y_+, t_+$) outward. Note that one of the key assumptions here is that the envelope is massive enough so that we can integrate outward in $q$ without reaching the photosphere (up to 30\% of the core mass in Fig. \ref{fig:envolturas}).   Fig. \ref{fig:envolturas} shows the result of these envelope integrations
for $y_+\in(10^{-4}, 0.5)$, $\log |\mathbb{C}|\in(-1, 5)$ and for a
Kramers' opacity law ($a=1$ and $b=-4.5$). Similar results are
obtained for a classical Thomson electron scattering opacity ($a=0$
and $b=0$) and other values of $t_+\sim 1$ (see Appendix \ref{app:envelopes}).
As it is clear from Fig.  \ref{fig:envolturas} two main families of
solutions are possible. Solutions in black correspond to solutions in
which the radiative gradient increases as we move outward from the
burning shell. In most cases these envelopes become convective ($\nabla=0.4$).
Conversely, solutions in orange correspond to those cases in which
the radiative gradient drops extremely fast as we move outward (in
fact more than exponentially as can be seen in
Fig. \ref{fig:envolturas}). As shown in Fig.  \ref{fig:envolturas}
these solutions quickly become isothermal at a temperature close to
that of the shell.  This second family of solutions does not
correspond to physical stellar envelopes as they cannot satisfy
photospheric boundary conditions. Using Eddington's approximation we
know that $d\log T/d\tau|_{\rm ph} \simeq 3/16 $ and from a zero-order
integration for the pressure $d\log P/d\tau|_{\rm ph} \simeq 3/2 $,
and consequently the temperature gradient near a stellar photosphere
must be $\nabla_{\rm ph}\simeq 1/8$. Such condition cannot be satisfied
by the orange solutions that stay radiative throughout the whole integration as  $\nabla_{\rm rad}$ decreases monotonically, and for which $\nabla_{\rm rad}\ll 1$ very close to the core. Interestingly, the watershed between
both families of solutions corresponds to those envelopes that start at the burning shell with a temperature gradient of
$\nabla_+=(a+1)/(4-b)$ (shown as a blue horizontal line in Fig. \ref{fig:envolturas}.

To better understand this behavior it is instructive to look at the derivative of the temperature gradient. Using eqs. \ref{eq:adimensional}
and  \ref{eq:nabla}, it is easy to show that
\begin{equation}
  \frac{d\nabla_{\rm rad}}{d \log q}=\nabla_{\rm rad}\left[\left((a+1)+(b-4)\nabla_{\rm rad}\right)\frac{d\log y}{d\log q}-1\right]
  \label{eq:deri_nabla},
\end{equation}  
where due to hydrostatic equilibrium ${d\log y}/{d\log q}<0$. It is
now clear that if $\nabla_{\rm rad}\leq (a+1)/(4-b)$ at some point,
then eq. \ref{eq:deri_nabla} behaves locally as $\frac{d\nabla_{\rm
    rad}}{d \log q}=\mathbb{A}\times\nabla_{\rm rad}$ with
$\mathbb{A}<0$ and $\nabla_{\rm rad}$ drops exponentially. Then
solutions with $\nabla_{\rm rad}\leq (a+1)/(4-b)$ already at the
burning shell are not physical. Physically meaningful solutions are
then confined to values of $\nabla>(a+1)/(4-b)$. On the other hand,
due to convection, we know that the temperature gradient cannot exceed
the adiabatic value. Then, we have shown that outer envelope imposes a
very strict constraint on the value of the temperature gradient
in the envelope, and immediately above the burning shell
  ---$(a+1)/(4-b)<\nabla<0.4$.

Interestingly, as it can be guessed from Fig. \ref{fig:envolturas}, physical
solutions with $\nabla_+$ near the adiabatic value will correspond to
envelopes that turn convective very quickly (deep convection). We see
that, unless we are in the case of deep convection, with the bottom of
the convective envelope closer than $q\simeq 1.1$, the temperature
gradient stays close to the critical limit $\nabla^{\rm lim} =
(a+1)/(4-b)$. Those solutions with values of $\nabla_+$ near
$(a+1)/(4-b)$ will show relatively massive radiative regions on top of
the burning shell and can even stay radiative until they reach the
photosphere\footnote{It is worth noting that solutions with
  $\nabla_+>(a+1)/(4-b)$ are not forced to increase monotonically and
  can start to decrease depending on the local value of $\frac{d\log
    y}{d\log q}$. In fact, we know from integrations of photospheric
  conditions immediately below the photosphere $\nabla_{\rm rad}\simeq
  \tau/(\tau+2/3)$, where $\tau$ is Rosseland optical depth.}. In
addition, one shared property of all physical solutions is that they
show a significant decrease in temperature outside the burning shell
and a very strong decrease in pressure, in particular for those
solutions that start with $y_+\ll 1$. In Appendix \ref{app:deep_conv}
we show that if the convective zone reaches all the way to the
burning shell, then detailed models predict a completely different
behavior and a much smaller radius.

Most importantly, for typical opacity laws we have $\nabla^{\rm lim}=
0.25$ (Thomson electron scattering, $a=b=0$) and $\nabla^{\rm lim}=
0.2353$ (Kramers' opacity, $a=1$, $b=-4.5$). This means that the
value of $\nabla$ above the burning shell  varies less than a factor
of two throughout the envelope. This result will be very useful to
make simple estimations in the next sections.

We end this section noting that, although these conclusions are
strictly valid for an opacity that follows a power law, the
conclusions are quite general. In fact, for any opacity law we can
locally approximate it by a power law ($\kappa\simeq \kappa_0 P^{a'+1}
T^{b'-4}$). Due to the faster-than-exponential nature of
eq. \ref{eq:deri_nabla} if $\nabla$ drops, in some region, below the
local critical value then the envelope solutions will quickly
become unphysical. As a consequence, in envelopes with real opacity
laws $\nabla$ needs to stay between the local critical value
$\nabla^{\rm lim} = (a'+1)/(4-b')$ and the adiabatic value
$\nabla_{\rm ad}$. In particular, for real envelopes near the burning
shell we will have in most cases $\nabla_+ \simeq (a+1)/(4-b)$ and
only for deep convection $\nabla_+\simeq \nabla_{\rm ad}$.

\subsection{Temperature gradient inside the burning shell}\label{sec:inside}

Having shown in Section \ref{sec:aboveshell} that $
(a+1)/(4-b)\lesssim \nabla_+\lesssim 0.4$ we now turn to analyze how
$\nabla$ changes inside the burning shell where $l(r)$ is not constant
anymore (see Fig. \ref{fig:shell}).

From eqs. \ref{eq:hydro0} and \ref{eq:transport0} we see that inside the burning shell we have
\begin{equation}
  \frac{dT}{dP}= \frac{3}{16\pi\, a\, c\, G }\frac{\kappa l}{m \, T^3}.
\label{eq:dpdt}  
\end{equation}
Using again that the opacity
inside the shell follows $\kappa=\kappa_0 P^a T^b$ we can write
\begin{equation}
\frac{T^{4-b}}{P^{a+1}}\nabla\frac{1}{l}= \mathbb{K},
\label{eq:inside}
\end{equation}
where $\mathbb{K}>0$ is a constant.

Evaluating the left-hand side of eq. \ref{eq:inside} both at the upper
boundary and the center of the shell, we obtain
\begin{equation}
\nabla_+=\left(\frac{T_s}{T_+}\right)^{4-b}\left(\frac{P_+}{P_s}\right)^{a+1}\nabla_s \frac{L_c+L_s}{L_c+L_s/2}.
\label{eq:inside2}
\end{equation}
We know from the high sensitivity of the energy generation to
temperature that temperature changes only slightly inside the burning
shell $T_s-T_+=\Delta T= T_s/\nu$. Let us call $x=\Delta P/P_s$
($P_+=P_s(1-x)$), then $dP/P=(\nabla_s)^{-1} dT/T$ and we see that
$x=(\nabla_s\nu)^{-1}$. Note that, as $\Delta T/T_s\ll 1$ and $
(a+1)/(4-b)\lesssim \nabla_+\lesssim 0.4$ then also $x\ll 1$ for
typical values of the opacity. Then eq. \ref{eq:inside2} can be
written as
\begin{equation}
\nabla_+ = \left(1+\frac{1}{\nu}\right)^{4-b}\left(1-x\right)^{a+1}\frac{1}{x\nu} \frac{L_c+L_s}{L_c+L_s/2}.
\label{eq:inside3}
\end{equation}
Defining $F$ as the factor due to the luminosity
of the core, $F=(L_c+L_s)/(L_c+L_s/2)$  ($F=2$ for inert cores), we can write at first order in $1/\nu$
\begin{equation}
  \begin{aligned}
\nabla_+ = &   
  \left\{1+\left(\frac{4-b}{\nu}\right)\right.\\
  \times & \left. \left[1-\left(\frac{a+1}{4-b}\right)\frac{1}{\nabla_s}\right]
  + \mathcal{O}(1/\nu)^2  \right\} F \nabla_s.
   \end{aligned} 
  \label{eq:inside3}
\end{equation}
We see here that the dominant factor for $\nu\gg 1$ is due to the
change in the luminosity throughout the shell ($F$) and that at the
dominant order the value of $\nabla_s$ is
\begin{equation}
  \nabla_s\simeq \frac{\nabla_+}{F} =\left[\frac{1/2+L_c/L_s}{1+L_c/L_s}\right]\, \nabla_+
  \label{eq:nabla_s}
\end{equation}  
We see that the temperature gradient is changed by a relative large
factor ${F}^{-1}$ in the small region of the burning shell. In particular, for an inert core ($F=2$, $L_c\ll L_s$) and the value of $\nabla_s$ becomes half its value at the upper boundary. Like $\nabla_+$, $\nabla_s$ is also tightly constrained by the outer envelope.

If we now estimate $\nabla_{-}$, we see that the temperature gradient
at the bottom of the burning shell is
\begin{equation}
  \nabla_{-} \simeq \left[\frac{L_c}{L_s+L_c}\right] \nabla_+,
  \label{eq:nabla_menos}
\end{equation}  
where it becomes clear that in the case of the inert core $L_c\ll L_s$
the material becomes isothermal at the lower boundary of the burning
shell, as we know from numerical models of low-mass stars (see Appendix \ref{app:numerical_mantles}).

Knowing the value of $\nabla_s$ we can make some useful estimations of how pressure and density change across the burning shell.
Using that at the burning shell $dP/P\simeq (\nabla_s)^{-1} dT/T$, and using the equation of state of an ideal gas we get
\begin{equation}
  \begin{aligned}  
  P_+\simeq& P_- \exp\left[\frac{-2F}{\nu \nabla_+}\right],\\
  \rho_+\simeq &\rho_- \frac{\mu_{\rm env}}{\mu_{\rm core}}\exp\left[\frac{-2}{\nu}
    \left(\frac{F}{\nabla_+}-1\right)\right],
    \end{aligned}
\label{eq:estimacion_P_Rho}  
\end{equation}  
where $\mu_{\rm env}$ and $\mu_{\rm core}$ are the mean molecular weights of the envelope and core, respectively.

\subsection{The upper mantle and the drop in $P$, $\rho$ and $T$ above the burning shell}
\label{subsec:upper}
We now turn to analyze how the radiative region above the burning
shell (from now on, the upper mantle) imposes some tight constraints on
the thermodynamical quantities at the upper boundary of the burning
shell, and consequently on the burning shell itself.

Let us first note that due to hydrostatic equilibrium the drop in pressure and temperature immediately above the burning shell is
\begin{equation}
  \left. \frac{dP}{P}\right|_+= \frac{\bar{P_c}}{P_+} \frac{dm}{M_c},\ \ \
  \left. \frac{dT}{T}\right|_+= \frac{\bar{P_c}}{P_+} \nabla_+ \frac{dm}{M_c}.
  \label{eq:drop_m}
\end{equation}  
This implies that, as the core becomes more compact and the pressure
contrast between the upper boundary of the shell and the core
increases by orders of magnitude ($\bar{P_c}\gg P_+$), the drop in
pressure and temperature above the shell will become large even in a
region of almost negligible mass ($\Delta m/M_c=(m-M_c)/M_c \ll
1$). In the following we will show that, when this happens, very tight
constraints on the burning shell can be derived. Then we will show
that these constraints are approximately valid even at the very early
stages of shell burning (i.e. immediately after the end of the main
sequence).

One of the key results from Section \ref{sec:aboveshell} is the fact
that $\nabla$ is strictly constrained above the burning shell to
values $(a+1)/(4-b)<\nabla<0.4$ which in practice prevents $\nabla_+$
from varying more than a factor two in that region. If we now restrict
ourselves to a region of negligible mass above the burning shell
($\Delta m/M_c \ll 1$) we can approximate
\begin{equation}
  \frac{dT}{dr}\simeq -\frac{-\mu_{\rm env} G M_c \bar{\nabla}}{\Re} \frac{1}{r^2},
 \label{eq:dTdr_MR} 
\end{equation}  
where $\bar{\nabla}$ is typical mean value of $\nabla$ that fulfills
$(a+1)/(4-b)<\bar{\nabla}\leq 0.4$.  Integrating this expression downward from a
point ($r=R_0$) where our approximations are still valid we can obtain
\begin{equation}
  T\simeq \frac{G\,M_c \mu_{\rm env}}{r\, \Re} \bar{\nabla} 
  \frac{(1-r/R_0 )}{(1-T_0/T)}
  \label{eq:Tr_M_outer}
\end{equation}
We see that in those cases where $T$, $\rho$ and $P$ drop significantly in a massless region above the burning shell, this is when the core becomes compact in the sense that $\bar{P_c}\gg P_+$, then we can move outwards to a point $R_o\gg r$ and $T_0\ll T$ and the factors on the right-hand side of eq. \ref{eq:Tr_M_outer} become close to unity. We will see below, however, that even under not very extreme conditions, the factors on the right are of order one.

In particular,  eq. \ref{eq:Tr_M_outer} sets a tight relation for the temperature immediately above the burning shell,
\begin{equation}
  T_+\simeq \frac{G\,M_c \mu_{\rm env}}{R_s\, \Re} \bar{\nabla} 
  \frac{(1-R_s/R_0 )}{(1-T_0/T_+)}
  \label{eq:T+_MR-0}
\end{equation}
Again, here the adimensional factor $(1-R_s/R_{\rm 0})$ will approach unity fast, as the density contrast between the shell and the core increases, as when $\rho_+\ll \rho_c$ it is possible to integrate outward to large values of $R_0$ keeping the approximation $m\simeq M_c$. It will be shown later that even in the early stages of shell burning these adimensional factors are all close to unity. It is then useful to write
all the factors close to one as\footnote{It is worth noting that, while eq. \ref{eq:T+_MR} has similarities with equation 10.2 in \cite{2005slfh.book..149F} (when $\tau=0$) they rely on very different justifications. While the assumption of $\tau\simeq 0$ in \cite{2005slfh.book..149F} is based on the analysis of polytropic envelope integrations in the U-V plane with no discussion about the role of convection, eq. \ref{eq:T+_MR} is based on the analysis done in Section \ref{sec:aboveshell}. One of the key ingredients in the derivation of eq. \ref{eq:T+_MR} is the fact that the development of convection together with outer boundary conditions strongly constrain the value of $\nabla$. This highlights the role of convection in the formation of bright red giants and explains why the early models that neglected the existence of convection \citep{1952ApJ...116..463S} did not form luminous red giants.}
\begin{equation}
  T_+\simeq \frac{G\,M_c \mu_{\rm env}}{R_s\, \Re}\bar{\nabla}\zeta,
  \label{eq:T+_MR}
\end{equation}  
where $\zeta$ will be very close to one\footnote{It must be emphasized that the argument presented in the next sections does not require $\zeta=1$ but only that $\zeta$ is of order one, even if within a factor of a few. The important feature is that $\bar{\nabla}\zeta$ is tightly constrained, and eq. \ref{eq:T+_MR} sets an effective constraint between $T_+$, $M_c$ and $R_s$. In fact the case of $\zeta=1$ corresponds to the case when boundary terms can be neglected and $\rho$, $T$ and $P$ can be assumed to have a simple power-law dependence on $r$, leading to a polytropic relation among them \citep[see Chapter \S 19.2 in][]{2015sess.book.....M}.} and
$(a+1)/(4-b)<\bar{\nabla}\leq 0.4$. We will see below that when the
density and pressure drop by orders of magnitude in a massless region
above the burning shell, then $\bar{\nabla}\simeq (a+1)/(4-b)$.

It is useful to note that, when the factors on the right in eq. \ref{eq:Tr_M_outer} can be neglected we also have that
\begin{equation}
  \frac{d\rho}{dr}\simeq - \frac{\rho}{r}(1-\bar{\nabla})\frac{(1-T_0/T)}{(1-r/R_0 )}
   \label{eq:drhodr_outer}
\end{equation}  

To have a more quantitative idea of the implications in
eqs. \ref{eq:Tr_M_outer}, \ref{eq:T+_MR}, and \ref{eq:drhodr_outer} we
can assume that energy is transported by radiation in the region
immediately above the burning shell. Note that if energy is
transported by convection the situation is straightforward as then
$\nabla$ is completely constant and $\nabla=0.4$. Again, if we
restrict to a region where $\Delta m/M_c \ll 1$, we can replace
eq. \ref{eq:hydro0} into eq. \ref{eq:transport0} and obtain
\begin{equation}
  dT= \frac{3 L_\star}{16\pi\, a\, c\, G }\frac{\kappa }{M_c \, T^3}\, dP.
\label{eq:dp_vs_dt}  
\end{equation}
 Writing the opacity dependence close to the burning shell as
 $\kappa=\kappa_0 P^a T^b$ ($a\geq 0$, $b\leq 0$ under normal
 conditions), eq. \ref{eq:dp_vs_dt} can be integrated in the standard
 way \citep{2012sse..book.....K} from a point ($r=R_0$) where our
 approximation is still valid down to the burning shell
\begin{equation}
  (T_{\rm 0}^{4-b}-T^{4-b})= \frac{4-b}{a+1}\frac{3\kappa_0\, L_\star}{16\pi a c G M_c}
  (P_{\rm 0}^{a+1}-P^{a+1}).
\end{equation}
We can see now that the value of $\nabla$ inside the upper mantle is
\begin{equation}
  \nabla= \frac{(4-b)}{(a+1)} \frac{[1-(T_{\rm 0}/T)^{4-b}]}{
  [1-(P_{\rm 0}/P)^{a+1}]}.
\end{equation}
 Under these conditions, the temperature and density at the
 upper boundary of the burning shell must fulfill
\begin{equation}
  T_+^{4-b}=\frac{\Pi}{\Theta} \frac{4-b}{a+1}\frac{3\kappa_0\, L_\star}{16\pi a c G M_c}P_+^{a+1},
  \label{eq:OuterPT}
\end{equation}
with $\Theta=(1-(T_{\rm 0}/T_+)^{4-b})$ and $\Pi=(1-(P_{\rm 0}/P_+)^{a+1})$.
As before, as soon as $P_c\gg P_s$ pressure and temperature drop quickly above the burning shell and  $\Theta/\Pi\simeq 1$. The value of $\nabla$ at the upper boundary of the burning shell is then
\begin{equation}
 \nabla_+=\left. \frac{d \ln T}{d\ln P}\right|_{+}=\frac{a+1}{4-b}\frac{\Theta}{\Pi}.
  \label{eq:nabla+}
\end{equation}
Then, once $\Theta\simeq 1$ and $\Pi\simeq 1$ we see that typical values of $\bar{\nabla}\simeq \nabla_+\simeq (a+1)/(4-b)$ are 4/17 for a Kramers' opacity and 1/4 for Compton scattering.

Interestingly, the envelope integrations discussed in Section \ref{sec:aboveshell} show that this also applies to massive radiative envelopes for which we cannot approximate $m\simeq M_c$ through all the integration. This is because $\nabla$, due to eq. \ref{eq:deri_nabla}, needs to stay close to $\nabla^{\rm lim} = (a+1)/(4-b)$ to avoid a runaway increase in $\nabla$, which would turn the envelope convective, but also larger than this critical value to avoid a runaway drop in  $\nabla$, which would make it impossible to satisfy photospheric boundary conditions.

\label{subsec:upper}
As discussed above, when $\rho_+\ll \rho_c$ and $P_+\ll P_c$ we can
strictly demonstrate that $\Theta\simeq 1$, $\Pi\simeq 1$,
$\zeta\simeq 1$, $T_0\ll T_+$ and $R_s\ll R_0$ and all previous
expressions are very simple. In Appendix \ref{app:early} we show that
these approximations are still acceptable in the early stages of the
burning shell, when the pressure and density contrast are not that
large. Readers not inclined to these rough estimations can skip this
Appendix and wait for Section \ref{subsec:toy}, where the predictions
of these approximation will be compared, and validated, against full
stellar evolution models.

\subsection{Energy release at the burning shell} \label{sec:release}
A second relation between temperature gradient, temperature, density,
and shell luminosity can be derived from the fact that $L_s$ is the
consequence of the energy generated in the shell itself. From the energy
conservation equation
\begin{equation}
\frac{dl}{dm}=\epsilon,
\end{equation}  
we can estimate that $L_s\simeq \langle \epsilon \rangle \Delta m = \langle \epsilon \rangle \rho_s  4\pi {R_c}^2 \Delta r$. Noting that almost all energy is released in a region where $\delta T/T\simeq 2/\nu$ ($T_s/\nu$ corresponds to half a burning shell), we can use that $\Delta r\simeq -(dT/dr)^{-1}\, 2\,T_s/\nu$
\begin{equation}
  L_s= \langle \epsilon \rangle   4\pi {R_c}^2  \rho_s \frac{-T_s\,2}{\nu}\left(\frac{dT}{dr}\right)^{-1}.
\label{eq:shell2}  
\end{equation}
Given that energy transport at the burning shell happens through
radiation, we can estimate that the temperature gradient at the
peak of the H-burning shell, i.e. in the middle of the burning shell
where $l_{\rm mid-shell}=L_c+L_s/2= (L_s/2) (1+2\,L_c/L_s)$, also fulfills
\begin{equation}
\left. \frac{dT}{dr}\right|_{\rm shell} \approx \frac{-3}{16\pi a c}\frac{\kappa_s \rho_s}{{R_c}^2{T_s}^3} (L_s/2+L_c).
  \label{eq:shell1}
\end{equation}
Replacing $L_s$ from eq. \ref{eq:shell1} into eq. \ref{eq:shell2}, and noting that the mean energy release is about half its value at the peak\footnote{This is exact for a linear luminosity increase in the shell.}, $\langle \epsilon \rangle \approx \epsilon(\rho_s,T_s)/2 \approx \epsilon_0/2 {\rho_s}^t{T_s}^\nu$, 
and replacing that in eq. \ref{eq:transport0}  we have

\begin{equation}
  \begin{aligned}
  \left(\left.\frac{dT}{dr}\right|_{\rm mid-shell}\right)^2\simeq & \frac{3}{8 a c \nu}\left(\frac{\rho_s}{T_s}\right)^2 \\ & \kappa(\rho_s,T_s)\epsilon(\rho_s,T_s) \left( 1+\frac{2 L_c}{L_s}\right)
  \end{aligned}
\label{eq:burningshell}  
\end{equation}  
or
\begin{equation}
  \begin{aligned}
    \left(\left.\frac{dT}{dr}\right|_{\rm mid-shell}\right)^2 \simeq & \frac{3}{8 a c}\frac{\kappa_0 \epsilon_0 \Re^a}{\mu^a}\\ & \frac{{\rho_s}^{2+s+t}\,{T_s}^{a+b+\nu-2}}{\nu} \left( 1+\frac{2 L_c}{L_s}\right).
      \end{aligned}
  \label{eq:burningshell2}
\end{equation} 

\subsection{Constraints set by the burning shell on the thermodynamic variables}\label{sec:key}
Eqs. \ref{eq:burningshell} and \ref{eq:burningshell2} tell us how the
temperature gradient has to be, for the total energy of the
shell to be released in a region where temperature drops by $\sim 2
T_s/\nu$. In addition to eqs. \ref{eq:burningshell} and \ref{eq:burningshell2},
due to the radiative transport and hydrostatic equilibrium equations
we know that at location $r=R_s$ the temperature gradient must be such
that it transports the local luminosity of the star at that point. This is
\begin{equation}
  \begin{aligned}  
  \left.\frac{dT}{dr}\right|_{\rm mid-shell} = & \nabla_s \frac{T_s}{P_s}\left(\left.\frac{dP}{dT}\right|_{\rm mid-shell}\right)  \\ =&
  \frac{\nabla_+}{F}\times\left(\frac{-GM_c\mu_s}{{R_s}^2\Re}\right)
    \end{aligned}
  \label{eq:dTdt-transport-hydro}
\end{equation}

Equating the gradients $dT/dt({\rm mid-shell})$ in
eqs. \ref{eq:burningshell} and \ref{eq:dTdt-transport-hydro} we find
that the existence of the burning shell forces
\begin{equation}
\begin{aligned}   
&  \left(\frac{-GM_c\mu_s}{{R_s}^2\Re}\right)^2\frac{\left( 1+\frac{2 L_c}{L_s}\right)}{ \left( 1+\frac{L_c}{L_s}\right)^2}\times (\nabla_+)^2= \\ & \frac{3}{2 a c \nu}\left(\frac{\rho_s}{T_s}\right)^2\kappa(\rho_s,T_s)\epsilon(\rho_s,T_s)
\end{aligned}   
\label{eq:the-equation}  
\end{equation}  
where we have replaced the factor $1/F$ by its dependence on the
luminosity to emphasize the behavior of the equation in the case of
luminous shells $L_c/L_s\ll 1$. Eq. \ref{eq:the-equation} is the key
equation to understand the behavior of the stellar structure in the
presence of a burning shell\footnote{Eq. \ref{eq:the-equation} is closely related to equation 10.5 in \cite{2005slfh.book..149F}. However, here the weird dependence on the cube of the radius of the burning shell has been replaced by the more reasonable ${R_s}^2$, and the weird appearance of the radiation density constant has been removed. In addition, the explicit appearance of $\nabla_+$ highlights the role of the constraints imposed by the envelope. }.  We have seen that
$(a+1)/(4-b)<\nabla_+<\nabla_{\rm ad}$ and, as soon as their core
becomes dense $\nabla_+\simeq (a+1)/(4-b)$ (eq. \ref{eq:nabla+}).  The
first thing to note is that in the case of a dim shell $L_s\ll L_c$
this equation becomes meaningless (in fact) as it only implies that
$\epsilon_s\simeq 0$, which is a trivial result as nothing should
happen in that situation. As soon as the energy generated in the
burning shell becomes relevant ($L_s\sim L_c$)
eq. \ref{eq:the-equation} starts to put constraints on the values of
$T_s$ and $\rho_s$ as a function of the mass and radius of the core
($M_C$, $R_s$).  We see that the key consequence of the development of
a burning shell is the emergence of an additional constraint
(eq. \ref{eq:the-equation}), between $\rho_s$, $T_s$ and $R_s$ and
$M_c$.

To fully appreciate the power of eq. \ref{eq:the-equation} we need to
complement it with eq. \ref{eq:T+_MR}, a constraint that is always
present in stellar envelopes. Noting that $T_s= T_+(1+1/\nu)$ we have
\begin{equation}
  T_s\simeq \frac{G\,M_c \mu_{\rm env}}{R_s\, \Re}(1+1/\nu) \nabla_+.
  \label{eq:T_MR}
\end{equation}
where we have approximated $T_0\ll T_s$ and $R_s\ll R_0$, which according to
our previous discussion are good approximations under rather general
situations. 
This equation implies an additional constraint on $T_s$, and together
with eq. \ref{eq:the-equation} it implies that the values of $T_s$,
$\rho_s$ and $P_s$ are completely defined by the values of $M_c$ and
$R_s$. In fact eqs. \ref{eq:the-equation}, and \ref{eq:T_MR}
demonstrate the key hypothesis in the derivations of ``shell homology
relations'' \citep[see Chapter \S 33.2 in][]{2012sse..book.....K},
i.e. that $T_s (M_c,R_s)$, $\rho_s(M_c,R_s)$, and $P_s(M_c,R_s)$.

An alternative version of eq. \ref{eq:the-equation} can be obtained by using eq. \ref{eq:T_MR} on the left-hand side of eq. \ref{eq:the-equation} to obtain
\begin{equation}
\begin{aligned}   
 & \left(\frac{\mu_s}{\mu_e}\right)^2\frac{\left( 1+\frac{2 L_c}{L_s}\right)}{ \left( 1+\frac{L_c}{L_s}\right)^2}(1+1/\nu)^2= \\ &\frac{3}{2 a c \nu}\left(\frac{{\rho_s}^2{R_s}^2}{{T_s}^4}\right)\kappa(\rho_s,T_s)\epsilon(\rho_s,T_s).
\end{aligned}   
\label{eq:the-equation_II}  
\end{equation}
 To good approximation we have
\begin{equation}
\left(\frac{\mu_s}{\mu_e}\right)^2\simeq \frac{3}{2 a c \nu}\left(\frac{{\rho_s}^2{R_s}^2}{{T_s}^4}\right)\kappa(\rho_s,T_s)\epsilon(\rho_s,T_s).
\label{eq:the-equation_III}  
\end{equation}

In the next section we will test and validate all the expressions
derived here by building a toy model of a low-mass red giant and
comparing it to the predictions of detailed full stellar
models. Before this, let us summarize what we have found so far. In
Section \ref{sec:aboveshell} we have shown that the existence of an
envelope of nonnegligible mass, together with the photospheric
boundary conditions, imply that the temperature above the burning
shell must follow $(a+1)/(4-b)\lesssim\nabla_+\lesssim 0.4$, where
$(a+1)/(4-b)\simeq 0.23\textendash 0.25$. Then, on Section
\ref{sec:inside} we have seen that the high temperature sensitivity of
nuclear reactions and hydrostatic equilibrium can be used to show that
the temperature gradient inside the burning shell is $\nabla_s\simeq
\nabla_+/F$, where $F\simeq 2$ when the shell is much more luminous
than the core ($L_s\gg L_c$).  As shown in Section \ref{subsec:upper}
this implies, among other things, that the temperature at the shell
follows $T_s\propto M_c/R_s$.  Finally, the fact that the luminosity
generated at the burning shell must also be transported through the
burning shell leads to a tight constraint of $\rho_s$ and $T_s$ as a
function of $M_c$ and $R_s$. Together these last two results
demonstrate the validity of the key hypothesis of the so-called shell
homology relations \citep[see Chapter \S 33.2
  in][]{2012sse..book.....K}, i.e. that we can write
$T_s(M_c,R_s)$, $\rho_s(M_c,R_s)$, and $P_s(M_c,R_s)$.

\section{The validation: A toy model for a low-mass star}\label{sec:toy-lowmass}
\begin{figure}
\centering
\includegraphics[width=\columnwidth]{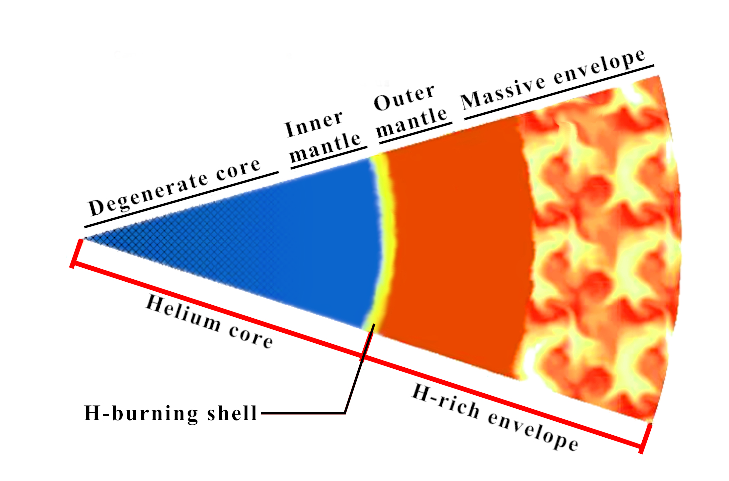}
\caption{Main structural parts of a low-mass red giant.}
\label{fig:Structure}
\end{figure}
To better understand the equations derived in Section \ref{sec:shell}
and to check their accuracy, we will construct here a toy model of a
low-mass red giant (Fig. \ref{fig:Structure}). Fortunately, the
degenerate core of low-mass red giants can be approximated by a
self-gravitating sphere sustained by the pressure of cold
nonrelativistic electrons.

A cold core supported by fully degenerate nonrelativistic electrons
has a very tight mass-radius relationship that corresponds to an
$n=3/2$ polytrope. In cgs units, this relationship is
\begin{equation}
  R_{\rm dc}=1.12\times 10^{20} {M_{\rm dc}}^{-1/3}.
 \label{eq:MR_dc}
\end{equation}
This tight constraint indicates that the geometrical size of the core
of low-mass stars only evolves due to the increase in their mass. This
increase in $M_c$ is provided by the steady burning of H at the
burning shell. Consequently the radius of the degenerate core will
evolve on a nuclear timescale. Then, due to the degeneracy of the
He core, gravothermal energy release is to a first approximation
negligible in comparison with the intensity of the H-burning shell;
i.e. mathematically we can say that $L_s \gg L_c$.  This fact,
together with the high thermal conductivity provided by degenerate
electrons will allow us to approximate the degenerate core as an
isothermal, thermally relaxed ($\int_0^{M_c} \epsilon_g dm=0$)
structure with $L_c\ll L_s$.

One feature, usually neglected when discussing low-mass giants, and
which is key to understand the $M_c(R_s)$ relation of low-mass red
giants, is the existence of a very light mantle of nondegenerate ideal
gas between the burning shell and the degenerate core. The existence
of such an inner mantle can be easily motivated as follows. Due to the
presence of the H-burning shell the stellar plasma at the shell cannot
be degenerate\footnote{Otherwise a flash would develop, injecting huge
  amounts of energy in that layer, and lifting degeneracy.}. Then, as
the material on the burning shell is not degenerate, and $T$, $\rho$,
and $P$ are continuous, there {\it must} exist regions immediately
below the burning shell where the material behaves as a classical
ideal gas ($P=\Re\, \rho\, T/\mu$). We will show that it plays a key
role in the $M_c(R_s)$ relation of low-mass red giants even when its
mass ($\Delta m$) is negligible (in the sense that $\Delta m \ll
M_c$).

In low-mass stars after the end of core H burning the contraction of
the tiny He core left is stopped by degeneracy, and the temperatures
around the H-exhausted core reach $T_s\sim 10^7$K. Note that this
happens before the development of a giant structure. From this point
onward, the star can be described as composed of an almost pure inert
He core, initially of less that a tenth of a solar mass and increasing
with time, and a H-rich envelope that harbors the rest of the mass of
the star, both separated by the presence of a H-burning shell. In
addition the He core harbors a He mantle where the material behaves as
a classical ideal gas. Such a configuration is depicted in
Fig. \ref{fig:Structure}.

\subsection{The inner mantle}
\label{sec:innermantle}

As evolution proceeds in a nuclear timescale we can assume at first order
that $L_c = 0$.  The temperature profile of the inner mantle is then
given by eq. \ref{eq:nabla_menos}, which for $L_c = 0$ corresponds to
$\nabla_-=0$.  We see that the presence of the H-burning shell forces
the existence of an isothermal mantle with $T\simeq
T_{-}=T_s(1+1/\nu)$ in between the H-burning shell and the degenerate
core\footnote{This mantle is clearly seen in detailed stellar models;
  see Appendix \ref{app:numerical_mantles}.}. This mantle is in
hydrostatic equilibrium and has to fulfill eq.  \ref{eq:hydro0}. As
long as the density does not increase significantly the isothermal
mantle can be considered as an ideal classical gas.  The mean
molecular weight of the inner mantle corresponds to that of pure and
fully ionized He ($\mu_c\simeq 4/3$). In this region the density must
follow
\begin{equation}
  \frac{d\rho}{dr}=-\frac{G m \mu_c}{\Re\, r^2 T_{-}}\rho,
  \label{eq:in_rho}
  \end{equation}
and if we restrict ourselves to the outermost regions of the core, where $\Delta m(r)\ll M_c$, the equation becomes
\begin{equation}
  \frac{d\rho}{\rho}=-\frac{G M_c \mu_c}{\Re\, T_{-}} \frac{dr}{r^2}.
  \label{eq:inner_rho_II}
  \end{equation}
This equation can be integrated inward from the burning shell
$r=R_s$, and will be a good description of the structure as long as
the electrons remain nondegenerate and the mass of the region is
negligible in comparison to that of the core ($\Delta m(r)\ll M_c$). As
$\rho_s$ is, in most cases, much lower than the mean density of the core
($\rho_s\ll \bar{\rho_c}$), then eq. \ref{eq:inner_rho_II} can
be integrated quite far from the burning shell ($|r-R_s|\lesssim R_s$)
and still fulfill $\Delta m(r)\ll M_c$. Integrating the density
inward we find that in the (almost) massless mantle below the burning
shell, the density follows
\begin{equation}
  \rho(r)=\rho_{-} e^{\frac{G \mu_c M_c}{\Re T_{-}}\left(\frac{R_s-r}{R_s\, r}\right)}.
  \label{eq:Inner_rho}
\end{equation}

Eq. \ref{eq:Inner_rho} is valid as long as $\Delta m(r)\ll M_c$, and when
$\rho_s\ll\bar{\rho_c}$ this equation will be valid far from
the burning shell $|r-R_s|\lesssim R_s$.

A cold core supported by fully degenerate nonrelativistic electrons
is described by eq. \ref{eq:MR_dc}. However, due to
eq. \ref{eq:Inner_rho} the density drops exponentially between the
degenerate core and the burning shell. The point at which the core
stops to behave as a degenerate electron gas corresponds to the point
at which the pressure from the nondegenerate ion gas starts to be
more important than that of the degenerate electrons. The density at
the boundary of the degenerate core ($\rho_{\rm Bdc}$) can be
estimated by setting $P_{e^-}=P_{\rm ions}$,\footnote{Note that it is
  reasonable to assume that electrons are degenerate in this
  estimation as the density at which electrons stop to behave as a
  degenerate gas corresponds to $\epsilon_{\rm Fermi}\approx kT$,
  which corresponds to
  $$
  \rho(\epsilon_{\rm Fermi}\approx kT)\approx 1.2\times 10^{-8}\, T^{3/2}
  $$
which is lower than the density in eq. \ref{eq:rho_Bdc}.}
\begin{equation}
  \rho_{\rm Bdc}\approx 1.7\times 10^{-8}\, {T_s}^{3/2} [cgs].
  \label{eq:rho_Bdc}
\end{equation}  
Where we have used that in the absence of neutrino cooling the
degenerate core behaves isothermally due to the high heat transport
efficiency of degenerate electrons. Note that for typical H-burning
temperatures the density at the boundary of the degenerate core
$\rho_{\rm Bdc}\ll \langle \rho_{\rm dc}\rangle =3 M_{\rm dc}/(4\pi
{R_{\rm dc}}^3)$, which means that $\rho_{\rm Bdc}$ is almost zero in
comparison with $\langle \rho_{\rm dc}\rangle $ and its location can
be approximated by radius of the $n=3/2$ polytrope
(eq. \ref{eq:MR_dc}). The condition $\rho_{\rm Bdc}\ll \langle
\rho_{\rm dc}\rangle $ also implies that eq. \ref{eq:Inner_rho} is
valid for values of $|r-R_s|\gtrsim R_s$, and the inner mantle can
 be treated as massless; $M_{\rm dc}\simeq M_c$. In particular,
if we now integrate the inner mantle (eq. \ref{eq:Inner_rho}) down to
the location of the upper boundary of the degenerate core, we get
\begin{equation}
   1.7\times 10^{-8}\, {T_s}^{3/2}\approx \rho_{\rm Bdc}\approx \rho_{-} e^{\frac{G \mu_c M_c}{\Re T_s}\left(\frac{R_s-R_{\rm dc}}{R_s\,R_{\rm dc} }\right)}.
  \label{eq:innermantle}
  \end{equation}
Eq. \ref{eq:innermantle} gives a new relation between $\rho_s$, $T_s$
and $M_c$ and $R_s$. This relation must be fulfilled together with
those coming from eqs.  \ref{eq:the-equation}, and \ref{eq:T_MR}. Note
in passing that each of these equations represents a relation imposed
by the three distinctive regions of our star: the core
(eq. \ref{eq:innermantle}), the burning shell
(eq. \ref{eq:the-equation}), and the envelope (eqs. \ref{eq:nabla+}
and \ref{eq:T_MR}).

\subsection{A complete toy}
\label{subsec:toy}
Before continuing with our analysis we will write
eq. \ref{eq:innermantle} in a more favorable way. Replacing
eq. \ref{eq:T_MR} into equation \ref{eq:innermantle}, and using $\nabla_+\simeq (a+1)/(4-b)$ we can write
\begin{equation}
\begin{aligned}   
  \rho_s \simeq  & 5.33\times 10^5 \frac{\mu_s}{\mu_c}
  \exp\left(-\frac{1}{\nu}\frac{(7-2b-a)}{(a+1)}\right)
\left(1+1/\nu\right)^{3/2} \\ &
  {T_9}^{3/2}  \,
  \exp\left(\frac{(4-b)}{(a+1)}\frac{1}{(1+2/\nu)}\frac{\mu_c}{\mu_{\rm env}}\left(1-\frac{R_{\rm dc}}{R_s}\right)\right).
 \end{aligned}  
   \label{eq:density}
\end{equation}

At the densities and temperatures typical of the burning shell and the
upper mantle, the opacity is dominated by electron
scattering\footnote{While the core is below $M_c\lesssim 0.23 M_\odot$
and the shell temperature is close to $\simeq 3\times 10^7$K, Thomson
scattering is not the only relevant source of opacity, and other
sources, mostly bound-free opacities are also important. It is only
after $M_c\simeq 0.25 M_\odot$ that Thomson scattering becomes dominant
at the shell temperature. In Appendix \ref{app:opac} we show that
qualitative behavior is unaltered when a Kramers' bound-free opacity law is adopted.}, and can be approximated as
\begin{equation}
\kappa_e=0.2(1+X)\,\, \hbox{\rm cm$^2$/g}.
\end{equation}
At the burning shell (hydrogen mass fraction $X\simeq 0.35$) we have
$\kappa_s \simeq 0.27$ cm$^2$/g. The energy generation due to CNO burning in that temperature range, $T_s/10^9=T_9\in(0.001, 0.1)$, is well approximated by
\begin{equation}
\epsilon_{\rm CNO}\simeq 8.24\times 10^{25} X_{\rm CNO} X \rho_s {T_9}^{-2/3}\exp\left({-15.231 T^{-1/3}}\right).
\end{equation}
Where for a typical composition  we have $ X_{\rm CNO}\simeq 0.01$. The choice of Thomson scattering and CNO burning corresponds to $a=b=0$ and $t=1$ in the expressions of $\epsilon$ and $\kappa$, as discussed in Section \ref{sec:shell}. The temperature dependence $\nu$ of the nuclear generation rate is, 
\begin{equation}
\nu=\frac{d\ln \epsilon_{\rm CNO}}{d\ln T_9}\simeq \left(-\frac{2}{3}+5.077\, {T_9}^{-1/3}\right).
\end{equation}  
This expression gives $\nu\simeq 23$, 13 and 10 for $T_9\simeq 0.01$, 0.05, and 0.1 respectively, as it should.

In these conditions, eq. \ref{eq:the-equation_III} becomes
\begin{equation}
  \left(\frac{\mu_s}{\mu_{\rm env}}\right)^2\simeq 5.099\times 10^{-10}
  \frac{{\rho_s}^3{R_s}^2{T_9}^{-14/3}}{\nu}\exp\left({-15.231\, {T_9}^{-1/3}}\right).
 \label{eq:rho-R-T} 
\end{equation}  
Replacing eq. \ref{eq:density} into eq. \ref{eq:rho-R-T} we have, for the case of Thomson scattering
\begin{equation}
\begin{aligned}   
  1.2953\times 10^{-8} & \frac{\mu_c^3}{\mu_s\mu_{\rm env}^2}\nu \exp(21/\nu)\left(1+1/\nu\right)^{-9/2} \\
  = & {T_9}^{-1/6} {R_s}^2  \exp\left[\frac{12}{(1+2/\nu)}\frac{\mu_c}{\mu_{\rm env}}\left(1-\frac{R_s}{R_{\rm dc}}\right) \right.\\
    - &\left. 15.231 {T_9}^{-1/3}\right]
 \end{aligned}  
 \label{eq:RT} 
\end{equation}
 
Now we can rewrite eq. \ref{eq:RT} and eq. \ref{eq:T_MR} in a more
convenient way by properly normalizing $R_s$, leaving an explicit
dependence on the mass of the core. These equations result in
\begin{equation}
  \frac{R_s}{R_{\rm dc}}\simeq \frac{\mu_{\rm env}}{\mu_c}\frac{0.6}{T_9}
  \left(\frac{M_{\rm dc}}{M_\odot}\right)^{4/3},
  \label{eq:xmT9}
\end{equation}
and
\begin{equation}
  \begin{aligned} 
 1.633\times 10^{-26} &\frac{\mu_c^3}{\mu_s\mu_{\rm env}^2}\nu \exp(21/\nu)\left(1+1/\nu\right)^{-9/2} \\ = &{T_9}^{-1/6} \left[\frac{R_s}{R_{\rm dc}}\right]^2\left[\frac{M_c}{M_\odot}\right]^{-2/3} \\ 
 \times & \exp\left[\frac{12}{(1+2/\nu)}\frac{\mu_c}{\mu_{\rm env}}\left(1-\frac{R_{\rm s}}{R_{\rm dc}}\right) \right. \\
   -&\left.15.231 {T_9}^{-1/3}\right]
 \end{aligned} 
 \label{eq:xT9} 
\end{equation}  
Typical values for the mean molecular weight are $\mu_c/\mu_s\simeq
1.58$, $\mu_c/\mu_{\rm env}\simeq 2.167$. The actual value of
$T_9(M_c)$ and $R_s(M_c)$ is not strongly affected by factors of order
1 on the left-hand side of eq. \ref{eq:xT9} and is dominated by the
exponents in $T_9$, $R_s$ and the exponential function. In particular,
we see that the exponential contains a factor $\mu_c/\mu_{\rm env}$
which explains the well-known fact that the temperature of the burning
shell, and consequently the luminosity of the shell, is affected by
the molecular weight contrast between the envelope and the core. This
is the reason why some previous works have found that molecular weight
contrast between the envelope and the core helps the development of
the giant structure \citep{2009PASA...26..203S}. In addition, these
expressions show us that estimating the impact of the molecular weight
on the luminosity of the shell by means of shell homology relations
where one considers the changes in $\mu$ and $R_c$ as independent is
not completely right \citep[see Chapter \S 33.2
  in][]{2012sse..book.....K}. In a full stellar model altering the
molecular weight of the envelope will have an impact on the radius of
the burning shell, which also alters the temperature of the shell.

\begin{figure}
\centering
\includegraphics[width=\columnwidth]{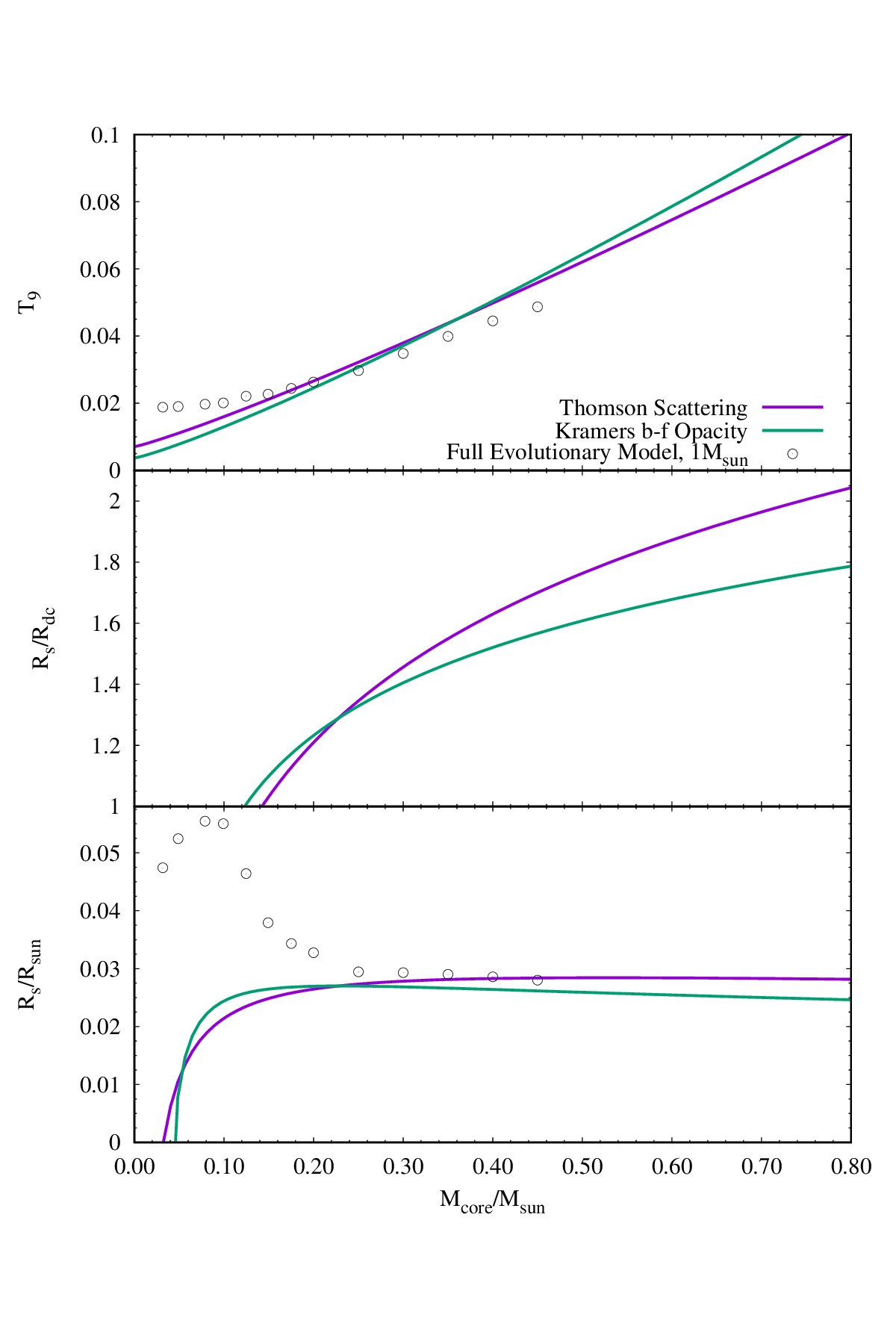}
\caption{Values of $R_s$ and $T_9$ for different core masses that fulfill eqs.\ref{eq:xmT9} and eq. \ref{eq:xT9} and the corresponding equations for Kramers' bound-free opacities; see Appendix \ref{app:opac}.}
\label{fig:RT}
\end{figure}

\begin{figure}
\centering
\includegraphics[width=\columnwidth]{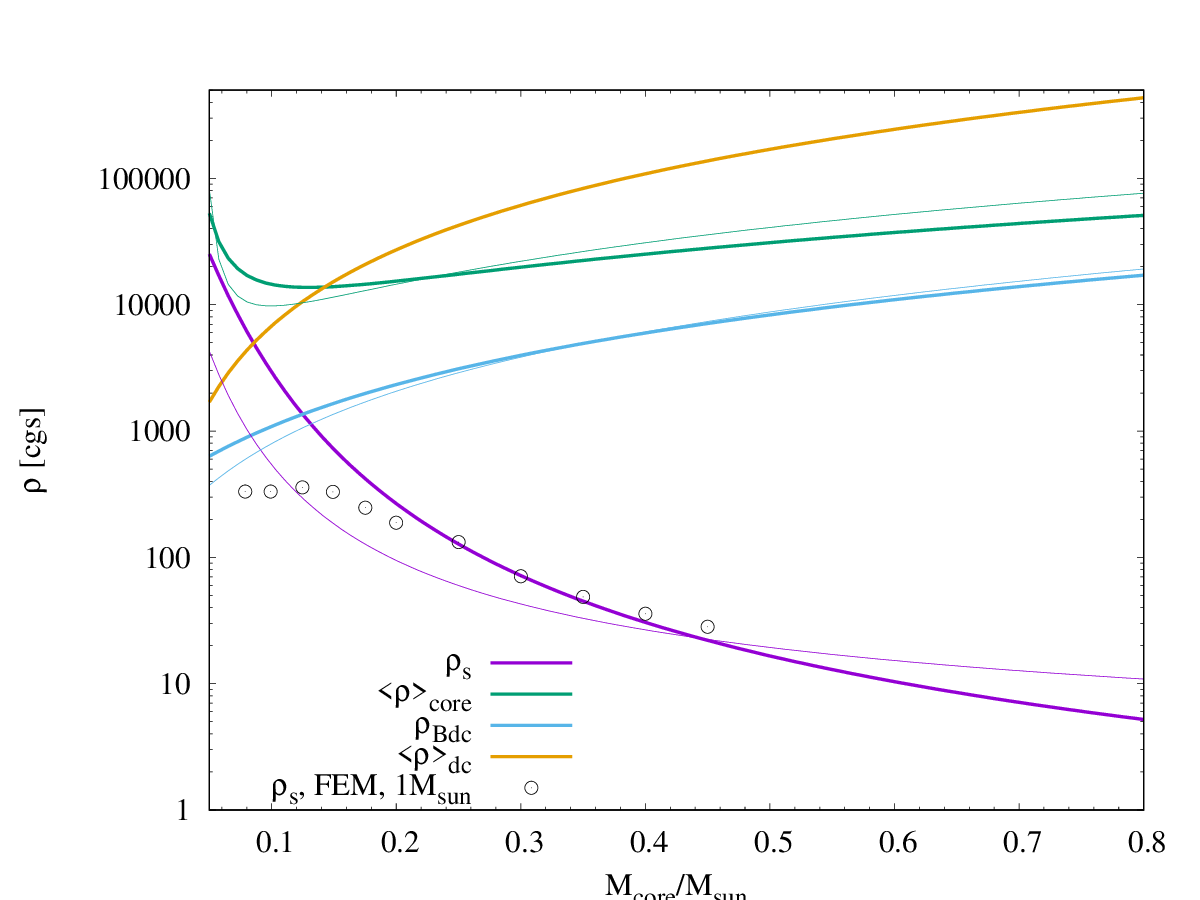}
\caption{Dependence of the density of the shell as a function of the mass of the core compared with other key densities of our  toy model. Thick lines indicate the relations obtained for Thomson scattering, while thin lines depict the same relations for a Kramers' bound-free opacity.}
\label{fig:densities}
\end{figure}

By replacing $T_9(R_s)$ or $R_s(T_9)$ from eq. \ref{eq:xmT9} into eq.  \ref{eq:xT9} we obtain the solutions for $T_9(M_c)$ and $R_s(M_c)$. These solutions are shown in Fig. \ref{fig:RT}, and can be very well approximated by
\begin{eqnarray}
  T_9&=&0.0071+0.12 \left(\frac{M_c}{M_\odot}\right)^{1.13}\nonumber\\
  \frac{R_s}{R_{\rm dc}}&=&2.181+0.589 \,\ln \left(\frac{M_c}{M_\odot}-0.008\right).
 \label{eq:TR_M-lowmass} 
\end{eqnarray}
Similar expressions for a Kramers' opacity are derived in
Appendix \ref{app:opac}. Fig. \ref{fig:RT} shows the predicted evolution of the
temperature and radius of the H-burning shell as the mass of the core
increases as a consequence of nuclear burning in our toy model.  The
resulting density of the burning shell is shown in
Fig. \ref{fig:densities}

We see that the burning shell settles near the degenerate core but
moves away from it as the shell becomes hotter. As a consequence the
burning shell settles at an almost constant value of $R_s\simeq 0.026$
(middle and lower panels in Fig. \ref{fig:RT}). The fact that our toy
model is able to reproduce the radius of the burning shell predicted
by the full stellar evolution model to less than 10\% is a remarkable
confirmation of the validity of our assumptions. We now see that the
constancy in the radius of the burning shell is the consequence of the
existence of an ideal gas mantle on top of a degenerate core. As the
mass of the core increases, it forces a drop in the density of the
burning shell. As the degenerate part of the helium core
  contracts (following $R_{\rm dc}\propto M_c^{-1/3}$) the density of the
  isothermal mantle between the degenerate core and the burning shell
  drops (so that density in this region can go from $\rho_{\rm Bdc}$ to
  $\rho_{s}$). This stretches the isothermal mantle of ideal gas, counterbalancing the mild contraction of the degenerate part of the core, and leading to an almost constant radius of the burning shell.

In spite of all simplifications in the treatment of the core and the
fact that the envelope is only required to be massive enough and
fulfill outer boundary conditions, the model gives a very good
quantitative agreement with the temperature and radius of the shell
derived from the detailed computation of the full set of stellar
evolution equations by means of a state-of-the-art stellar evolution
code \citep[{\tt LPCODE},][]{2016A&A...588A..25M}. This is
particularly true once the thin shell approximation adopted throughout
this paper becomes good and the density contrast between the shell and
the core is high, but the agreement is still good even at the
beginning of the shell burning stage as $R_s$ in the simple model deviates by less than a factor two from the real value of $R_s$
(see Fig. \ref{fig:RT}).  This level of agreement is an astonishing
confirmation of the accuracy of eqs.  \ref{eq:nabla+},
\ref{eq:T_MR} and \ref{eq:the-equation_III} and the approximations
$\Theta\simeq 1$, $\Pi\simeq 1$, $\zeta\simeq 1$ and $R_s\ll R_0$ on
which it is based. As we mentioned at the beginning of this article,
toy models are not made to be precise but to be an accurate
description of the main mechanisms of a given process, so we find this
level of precision remarkable.

Besides the confirmation of our key equations, our toy model offers
some very interesting insight on the structure of low-mass red
giants. The model shows why the existence of a mantle of negligible
mass above the degenerate core increases its radius, so that, as the
core increases its mass and the radius of the degenerate regions
decreases as ${M_c}^{-1/3}$, the radius of the burning shell stays
at a constant value of $R_s\simeq 0.027... 0.03 R_\odot$.  In
addition, the model helps to understand why all low-mass stars develop
a He flash when the core mass is about half a solar mass irrespective
of initial mass, metallicity, or mass of the envelope.  We see that,
for a star with a completely isothermal and thermally relaxed core,
the H-burning shell (and consequently the core) reaches He-burning
temperatures when the mass of the core is $M_c\simeq 0.75 M_\odot$,
independent of the metallicity and the mass of the envelope. This sets
an upper limit for the mass of the core at the moment of the He-core
flash.  The fact that the He flash happens before the core reaches
this limiting mass is due to gravothermal heat release (see discussion
in Appendix \ref{app:numerical_mantles}). As the core increases its
mass and the H-burning shell heats up, the timescale of nuclear
reactions becomes shorter and comparable with the thermal
(Kelvin-Helmholtz) time scale of the the core.  This additional
heating of the core anticipates the development of the
He-flash. However, as this additional heating is a function of the
rate at which mass is added to the core ($\dot{M_c}$), and $\dot{M_c}$
is solely determined by $T_s$, the mass of the core at which the
helium flash will happen will be almost constant and irrespective of
initial mass, metallicity, and mass of the envelope.

\section{Dimensional analysis}\label{sec:dym}

Now that we have confirmed the accuracy of the equations derived in
Section \ref{sec:shell} we can turn to analyze their implications
for the global structure of stars.

From eq. \ref{eq:T_MR} we know that $T_s\propto M_c/R_s$, replacing now this into eq. \ref{eq:the-equation} we find that
\begin{equation}
P_s= \mathcal{K}\, {R_s}^{(\nu-t-8+b)/(a+t+2)} {M_c}^{(6-\nu+t-b)/(a+t+2)},
\label{eq:dym_Ps}
\end{equation}
where $\mathcal{K}$ is a constant that can be computed from
eq. \ref{eq:the-equation}.  Due to hydrostatic equilibrium we know
that the mean pressure in the core is
\begin{equation}
  \bar{P_c}\approx G\,{M_c}^2/8\pi\,{R_s}^4.
\label{eq:dym_Pc}    
\end{equation}
Eqs. \ref{eq:dym_Ps} and \ref{eq:dym_Pc} show that the action of the
burning shell destroys the possibility of homologous contraction for
these stars as it gives
\begin{equation}
  \frac{P_s}{\bar{P_c}}=\mathcal{K'} {R_s}^{\frac{\nu+3t+b+4a}{a+t+2}}{M_c}^\frac{2-\nu-t-b-2a}{a+t+2}.
 \label{eq:dym_Ps-Pc}   
\end{equation}  
In fact as $\nu$ is the dominant term in both exponents, eq. \ref{eq:dym_Ps-Pc} tells us that, as soon as $R_s$ contracts or $M_c$
increases by a small amount, pressure at the burning shell drops by
orders of magnitude. To quantify this point lets focus in the most common situation in both low- and high-mass red giants, Thomson scattering ($a=b=0$) with CNO burning ($t=1$ and $\nu\simeq 23...13$). Under those conditions, eq. \ref{eq:dym_Ps-Pc} gives
\begin{equation}
  \frac{P_s}{\bar{P_c}}=\mathcal{K'} {R_s}^{1+\nu/3}{M_c}^\frac{1-\nu}{3}.
 \label{eq:dym_Ps-Pc}   
\end{equation} 
For typical values of $\nu$, $(1+\nu/3)\simeq 8.5 ... 5.3$, then a
small decrease in the radius of $\sim 10$\% leads to a $\sim
50... 80$\% drop in the ratio between the pressure at the shell and
that at the core. This is the situation in massive and
intermediate-mass stars (IM\&M), where the location of the shell
decreases by at least a factor of two, between the end of core H
burning and the beginning of He burning, while the mass of the core
remains constant, leading to a huge drop of more than two orders of
magnitude in $P_s/\bar{P_c}$. Note in passing that for IM\&M stars
$\dot{r}$ does not change sign at the maximum of energy release, as
stated by \cite{2012sse..book.....K}, but immediately above the
burning shell in the region where the density drops many orders of
magnitude.  On the contrary, in low-mass stars, as we have seen in the
previous section the interplay between the inner mantle and the
degenerate part of the core renders the position of the shell almost
constant (i.e. $\dot{r}=0$) while the mass of the core increases.

Using eq. \ref{eq:T_MR} and the equation of state we can also obtain
\begin{equation}
\rho_s= \mathcal{K''}\, {R_s}^{(\nu-6+b+a)/(a+t+2)} {M_c}^{(4-\nu-a-b)/(a+t+2)}.
\label{eq:dym_rhos}
\end{equation}

A knowledgeable reader will recognize the exponents in
eqs. \ref{eq:dym_Ps}, \ref{eq:dym_Ps-Pc} and \ref{eq:dym_rhos} as those provided by shell
homology relations. This should come as no surprise as in Section
\ref{sec:key} we mathematically demonstrated that to this level of
approximation the key hypothesis of shell homology relations is
rigorously valid (i.e. that $T_s (M_c,R_s)$, $\rho_s(M_c,R_s)$ and
$P_s(M_c,R_s)$). Consequently, any dimensional analysis based on our
relations should cast the same result. The main difference between
eqs. \ref{eq:dym_Ps} and \ref{eq:dym_rhos} and those derived by shell
homology relations is that here we are able to compute (if required)
the numerical values of the proportionality constants
$\mathcal{K}$,$\mathcal{K'}$, and $\mathcal{K''}$.

\section{A red giants' toy story}\label{sec:toystory}
We are now in condition to qualitatively understand the development of 
red giant structures after the main sequence.
The existence of an envelope of nonnegligible mass forces the mean
value of $\nabla\simeq 0.23 \textendash 0.4$ above the burning shell, which
leads to a tight dependence of the shell temperature on the mass and
radius of the core $T_s\propto M_c/R_s$
(eq. \ref{eq:T+_MR}). Additionally, the switching-on of the burning
shell and the fact that the heat is both generated and transported at
the burning shell leads to a tight relationship between $\rho_s$,
$T_s$, $M_c$ and $R_s$ (eq. \ref{eq:the-equation}). Together,
eqs. \ref{eq:T+_MR} and \ref{eq:the-equation} imply that $\rho_s(M_c,
R_s)$. For a given mass-radius relation of the core, $M_c(R_s)$ this implies
that the shell density will evolve as a function of $M_c$ or $R_s$.
Below we will analyze the typical mass-radius relations for low-mass
stars and IM\&M stars.  Interestingly we will see that
these two cases correspond quite well with evolutions at constant
$R_s$ or $M_c$, respectively.

The situation is simpler in more massive stars, where the core is an
undifferentiated contracting sphere with the same equation of state,
while it is more complicated for low-mass stars where the core can be
divided into a degenerate core and a surrounding mantle that behaves
as an ideal classical gas. In the latter case we will make use of the
simple $R_s(M_c)$ relation derived in Section \ref{sec:toy-lowmass}.

\subsection{Low-mass stars}\label{sec:toylow}

As soon as H burning is exhausted in the core the layers around it
start to contract and heat up as a consequence of the negative
gravothermal specific heat of typical of nondegenerate stellar
material\footnote{In stars with masses between $1.1M_\odot\lesssim
  M_\star \lesssim 2M_\odot$ the former convective core first
  contracts in a Kelvin-Helmholtz timescale until the layers
  surrounding the He core reach H-burning temperatures. On the
  contrary, in lower-mass stars $M_\star \lesssim 1.1M_\odot$ the
  transition from core H burning to shell H burning happens
  continuously.}, leading to the ignition of a H-burning shell around
the exhausted core.

The formation of a degenerate core surrounded by a hot mantle (Section
\ref{subsec:toy}) leads to stringent mass-radius relation for the
core.  Together with the constraints coming from the massive envelope
and the burning shell (eqs. \ref{eq:T+_MR} and \ref{eq:the-equation})
this forces the radius of the burning shell to settle at a very
specific constant location and at specific values of temperature and
density, all of which are solely defined by the mass of the core below
the burning shell (eq. \ref{eq:TR_M-lowmass} and Figs.  \ref{fig:RT}
and \ref{fig:densities}).  As $M_c$ only increases due to the
H consumption at the burning shell, $T_s(M_c)$,
$\rho_s(M_c)$,$P_s(M_c)$,$R_s(M_c)$ will evolve on a nuclear
timescale.  For low-mass red giants (eq. \ref{eq:TR_M-lowmass}) $R_s$
is forced to remain almost constant, and we get $P_s/\bar{P_c}\propto
{M_c}^{(2-\nu-t-b-2a)/(a+t+2)}$ and $\rho_s/\bar{\rho_c}\propto
{M_c}^{(2-\nu-2a-b-t)/(a+t+2)}$.  At first the switching-on of the
H-burning shell does not lead to a particularly increase in the
geometrical size of the star, because for low core masses $M_c\lesssim
0.12M_\odot$ the density of the shell does not differ so much from
that of the core, which means that similar amounts of mass are
harbored in similar volumes both in the core and around the burning
shell (Fig. \ref{fig:densities}). However as the mass of the core
increases due to nuclear burning, the density of the shell and the
density of the core evolve in opposite directions and soon much larger
volumes are needed to harbor masses around the burning shell. In our
toy homogeneous model already at $M_c\approx 0.12M_\odot$ the density
of the shell is about one order of magnitude lower than the the mean
density of the core $\bar{\rho_{\rm c}}$.  The drop in
density of the shell leads to a drop in the density of the envelope
leading to an increase in size. This situation is made extreme by the
enforcement of hydrostatic equilibrium on the upper mantle as we will
show in the next paragraph.

Eq. \ref{eq:drhodr_outer} is valid as long as $m(r)\simeq M_c$. Let us
define a point $R_{\rm BAP}$ which sets the boundary of the validity
for the massless approximation, i.e. $(m(R_{\rm
  BAP})-M_c)=\mathcal{Q}\, M_c$ with $\mathcal{Q}\ll 1$. For example,
with $\mathcal{Q}=0.1$ our estimations of $T$, $\rho$ in that mantle
will be wrong by less than 10\%. Within that upper mantle density drops as
\begin{equation}
  \rho(r)\approx \rho_s \left(\frac{r}{R_s}\right)^{-\delta}
\end{equation}  
with $\delta=(3-b-a)/(a+1)$. For the usual situation of Thomson
Scattering $\delta=3$. Integrating the mass up to $R_{\rm BAP}$ we have
\begin{equation}
  \Delta M= \mathcal{Q}\, M_c= \rho_s \int_{R_s}^{R_{\rm BAP}} 4\pi r^2
  \left(\frac{r}{R_s}\right)^{-\delta} r^2 dr.
\end{equation}
Solving the integral and using the definition of $\bar{\rho_c}$ we find that, for $\delta=3$
\begin{equation}
  R_{\rm BAP}= R_s \exp\left(\frac{\mathcal{Q}}{3}\frac{\bar{\rho_c}}{\rho_s}\right),
 \label{eq:RBAP} 
\end{equation}  
which is a remarkable result. Eq. \ref{eq:RBAP} shows that as soon as
the density contrast between the core and the shell increases the
radius of the \emph{massless} upper mantle on top of the shell
increases exponentially. This is a very well-known feature of
numerical models. The situation is even more extreme for bound-free opacities $\delta=3.5$ and density drops faster.

The direct consequence of this drop in the density contrast is
that the boundary of the massless mantle is moved far away from the
core ($R_{\rm BAP}\gg R_s$), and the density of the material drops
orders of magnitude to
\begin{equation}
  \rho_{\rm BAP}=\rho_s \exp\left(\frac{\mathcal{Q}}{3}\frac{\bar{\rho_c}}{\rho_s}\right)^{-\delta}.
  \label{eq:RhoBAP} 
\end{equation}  
while the pressure drops as
\begin{equation}
  P_{\rm BAP}=P_s \exp\left(\frac{\mathcal{Q}}{3}\frac{\bar{\rho_c}}{\rho_s}\right)^{-(1+\delta)}.
  \label{eq:P-BAP} 
\end{equation} 
As pointed out by \cite{2005slfh.book..149F},  the star is then left with the daunting task of harboring the whole
massive envelope with a very low density and under a very low
gravitational potential (due to the large increase in $R_{\rm BAP}$).
A simple estimation from the hydrostatic equilibrium equation (setting $P_{\rm surface}\approx 0$) gives
\begin{equation}
  \bar{R_{\rm env}}\simeq \frac{(G/4\pi)^{1/4}(M_c\, M_{\rm env})}{P_{\rm BAP}^{1/4}},
  \label{eq:Renv} 
\end{equation} 
and the exponential drop in $P_{\rm BAP}$ leads to a large increase
in the value of the typical radius $ \bar{R_{\rm env}}$ of the mass shells in
the massive envelope. Note that this qualitative conclusion is valid
regardless of  whether the massive envelope is radiative or
convective\footnote{It is affected, however, by very deep convection
  so that convection reaches down to the burning shell itself. In
  fact, very deep convection is known to reduce the size of red giant
  stars; see Appendix \ref{app:deep_conv}.}

\begin{figure}
\centering
\includegraphics[width=\columnwidth]{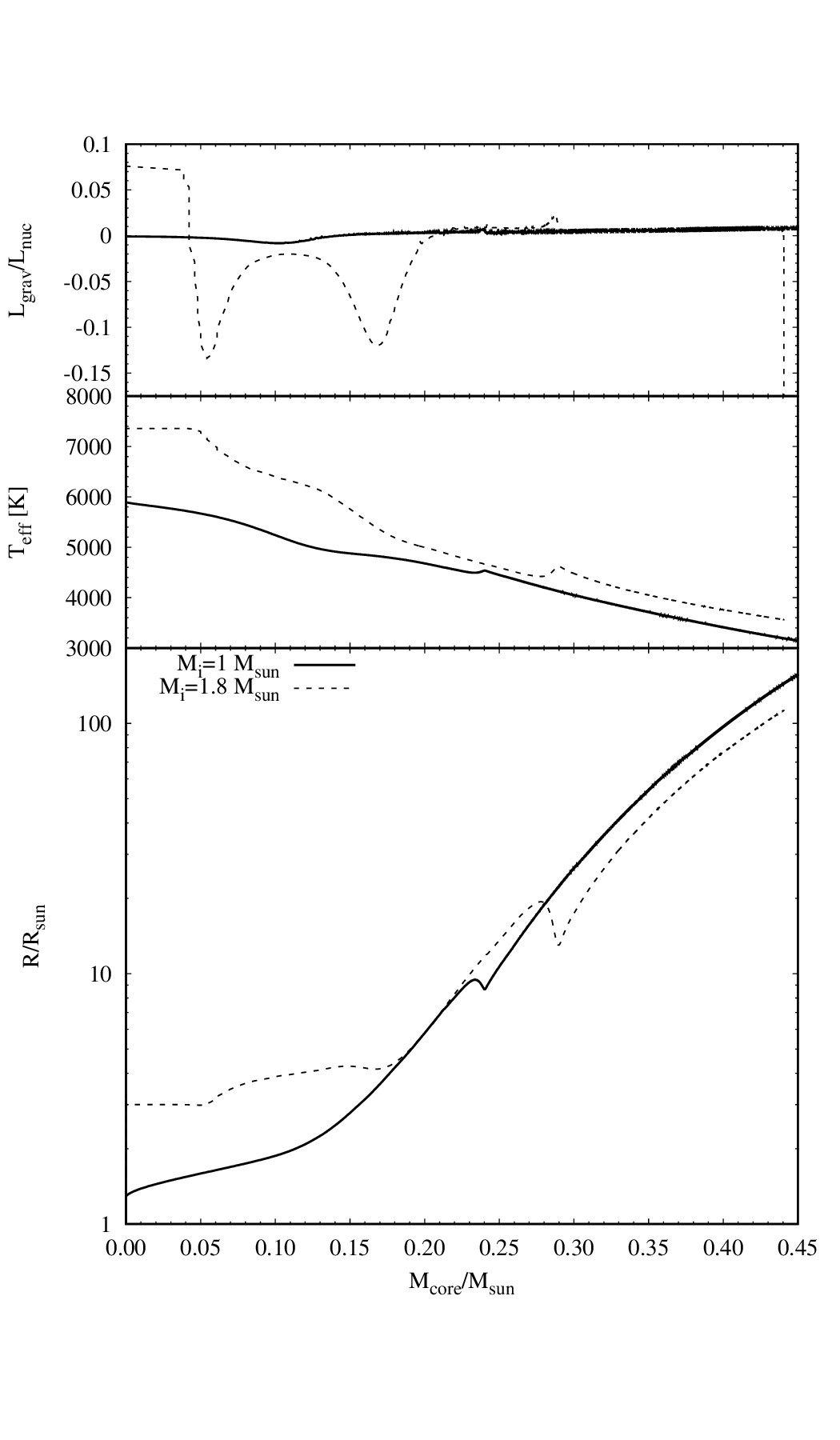}
\caption{Evolution of the radius of a $1 M_\odot$ model ($Z=0.02$) as a function of the mass of the He core.}
\label{fig:1Msun}
\end{figure}

Note that while $\rho_s$ stays within one order of magnitude of
$\bar{\rho_c}$ the exponent $(\mathcal{Q}\bar{\rho_c})/(\rho_s 3)$ is
close to unity and nothing happens. In our toy model
(Fig. \ref{fig:densities}) this corresponds to $M_c \lesssim 0.1
M_\odot$ (depending on the exact value of $\mathcal{Q}$ adopted). We
see that our simple model predicts that the change in the behavior of
the radius of low-mass red giants should appear once $M_c\gtrsim 0.1
M_\odot$, but not before. Fig. \ref{fig:1Msun} shows the behavior of
the stellar radius as a function of the mass of the core for  1 $M_\odot$ and 
1.8 $M_\odot$ full evolutionary models computed with {\tt LPCODE}. It
is clear from Fig. \ref{fig:1Msun} that there is a change in the
behavior of the radius as a function of the mass of the core at $M_c\sim
0.125 M_\odot$. This is particularly easy to see in the case of the $1
M_\odot$, in which H burned in the core radiatively and thus had a smooth
transition to shell burning. Stellar models with cores smaller than
this threshold change their radii by less than a factor of 2 even with an
increase in the mass of the core of more than one order of
magnitude.

\subsection{Intermediate-mass and massive stars (IM\&M)}
The situation in IM\&M stars is fortunately much
simpler that that of low-mass stars. This is because their core is not
differentiated and can be modeled as a contracting sphere of ideal
gas. For stars in this mass range the mass of the exhausted core after
the main sequence is larger\footnote{For stars below $\simeq 3 M_\odot$
  this limit is reached later after enough fresh He has been added to
  the core by the burning shell, see \cite{2005essp.book.....S}.} than
the Sch\"onberg-Chandrasekhar limit, and the exhausted core contracts
on a thermal timescale. As this timescale is shorter than the nuclear
timescale at which the burning shell adds mass to the core, the
evolution will proceed at almost constant $M_c$. As soon as core H
burning is finished and as temperatures around the burning shell reach
$T\simeq 10^7$K we can estimate the mean pressure in the core to be
\begin{equation}
  \bar{P_c}\approx \frac{G {M_c}^2}{8\pi {R_s}^4}+P_s,
  \label{eq:pcore}
\end{equation}  
and the mean density of the core is
\begin{equation}
\bar{\rho_c}\approx \frac{3 {M_c}}{4\pi {R_s}^3}.
\end{equation}  
As the equation of state in the contracting cores of massive and intermediate-mass stars corresponds to that of a classical ideal gas, we know that typical temperatures in the core are
\begin{equation}
  \bar{T_c}\approx \frac{\mu \bar{P_c}}{\Re\bar{\rho_c}}.
\end{equation}  
The evolution of those cores is ruled by the speed at which they lose
energy from their surface, as $L_c\neq 0$. The Virial theorem for such structures tell us that
\begin{equation}
2E_i+E_g =4\pi {R_s}^3 P_s
\end{equation}  
Using the expression for the internal energy of an ideal monoatomic gas
and for the gravitational energy we find that
\begin{equation}
k'\frac{3 M_c\Re \bar{T_c}}{\mu_c}-\frac{k G {M_c}^2}{R_c} =4\pi {R_s}^3 P_s
\end{equation}  
where $k'$ and $k$ are form factors of the order of unity that depend on
the details of the mass distribution of the core. In the most representative case of Thomson scattering, $P_s$ can be replaced as a function of $M_c$ and $R_s$ to give:
\begin{equation}
  k'\frac{3 \Re \bar{T_c}}{\mu_c}=\frac{k G {M_c}}{R_c} +4\pi k'' {R_s}^{\nu/3} M^{(4-\nu)/3}
  \label{eq:virial}
\end{equation}  

The time evolution of the core of massive stars is given by
$-d(E_i+E_g)/dt=L_c$, and consequently $L_c=dE_i/dt=-(dE_g/dt)/2$.  As
heat is radiated away from the core ($L_c>0$) at constant
mass\footnote{Because contraction happens on timescales much shorter
than those of nuclear reactions.}, the temperature will rise and the radius will
shrink. Due to the large value of $\nu$, the surface terms in
eqs. \ref{eq:virial} and \ref{eq:pcore} become quickly irrelevant.
As contraction proceeds at constant mass we see that
\begin{eqnarray}
  \frac{\rho_s}{\bar{\rho_c}}\propto {R_s}^{1+\nu/3} \nonumber\\
  \frac{P_s}{\bar{P_c}}\propto {R_s}^{1+\nu/3} \nonumber\\
   \frac{T_s}{\bar{T_c}}\propto {R_s}^{0}=1 \\
 \label{eq:evolucion}
\end{eqnarray}  
Due to the large exponents we see that the contrasts in the density and
pressure increase very rapidly with moderate contractions of the core.
Again, as in the case of low-mass stars, the development of a large
density contrast has enormous consequences for the hydrostatic
equilibrium of the upper mantle on top of the burning shell.  In fact,
the derivations of eqs. \ref{eq:RBAP}, \ref{eq:RhoBAP},
\ref{eq:P-BAP}, and \ref{eq:Renv} done in Section \ref{sec:toylow} are
independent from the specific characteristics of the core and are
valid also when the stellar core corresponds to an ideal classical
gas. A small contraction of the core leads to an increase in the
density contrast of the star and, as before, this will lead to a
further drop in the pressure and density ($P_{\rm BAP}$, $\rho_{\rm BAP}$)
and an increase in the radius ($R_{\rm BAP}$) at the bottom of the
massive envelope. The envelope of mass $M_{\rm env}$ has to be
accommodated in a very low gravitational potential and with very low
densities, which causes a huge increase in the typical radius
($\bar{R_{\rm env}}$) of its layers.

\begin{figure}
\centering
\includegraphics[width=\columnwidth]{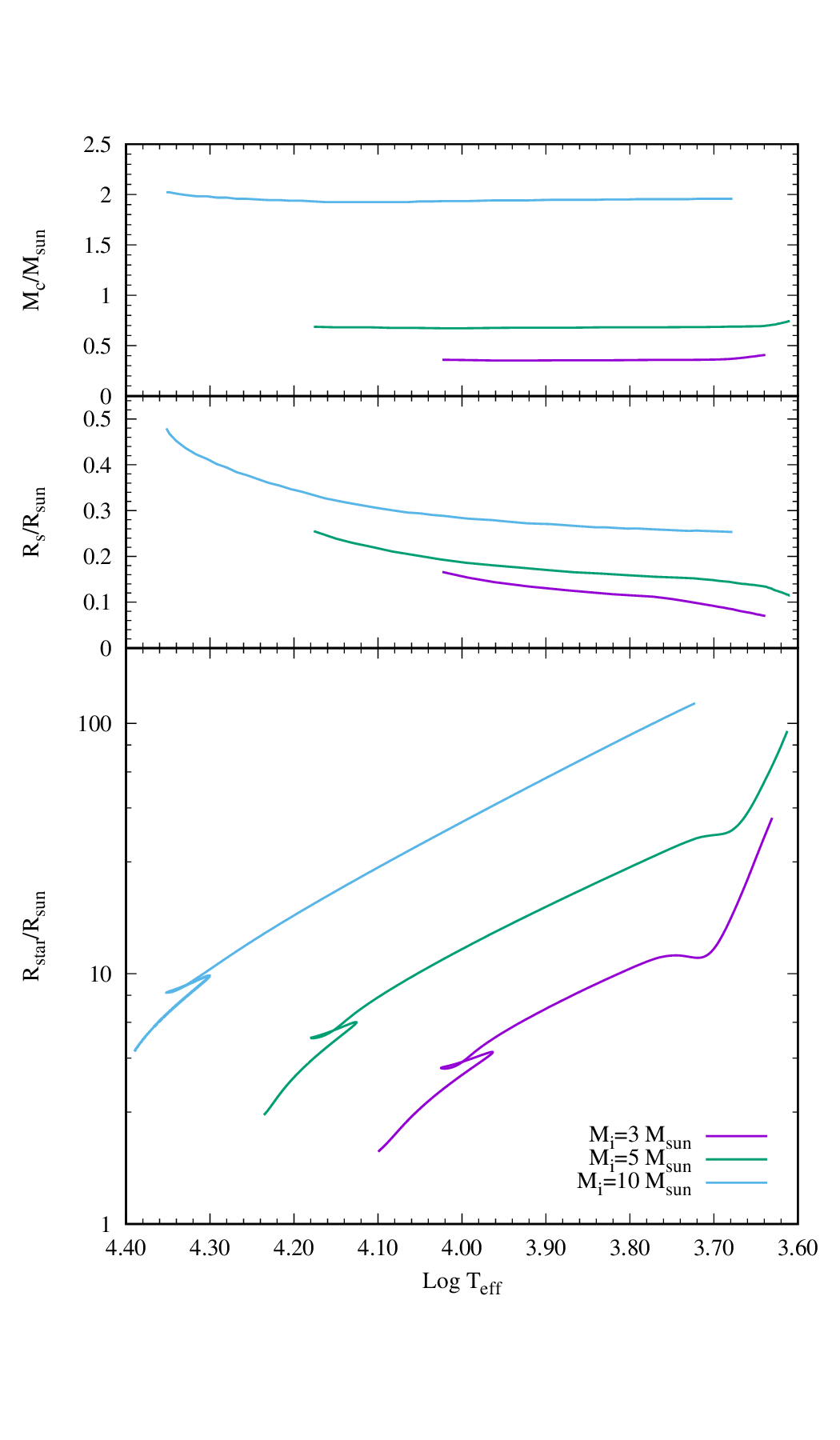}
\caption{Evolution of the core mass $M_c$, shell radius $R_s$ and
  stellar radius of three full evolutionary sequences with initial
  masses $3\ M_\odot$, $5\ M_\odot$, and $10\ M_\odot$ ($Z=0.02$).}
\label{fig:3-5-10Msun}
\end{figure}

Note that although the shell remains at an almost constant location in
the case of low-mass stars this is not true for IM\&M stars (see
Fig. \ref{fig:3-5-10Msun}). Due to the different central temperatures
at the end of the main sequence ($T/10^7=$ 3.6, 4.3, and 5.8 at the
end of the main sequence for the 3,5 and 10 $M_\odot$ models),
different degrees of core contraction are required before He-burning
temperatures are attained ($T_c\sim M_c/R_s$). For example, in the
sequences shown in Fig. \ref{fig:3-5-10Msun} between the end of the
main sequence and the beginning of He burning, the core contracts by
$R_s/R_s^0\simeq $2.5, 2.2 and 1.9 for the 3, 5 and 10 $M_\odot$
sequences. These changes in $R_s$ correspond to increases in the density
of $\bar{\rho_c}/\bar{\rho_c}^0\simeq 15.6$ 10.6, and 6.8.

\section{Conclusions and final remarks}\label{sec:end}

In this work we have revisited the long standing question of ``why
stars become red giants'' by means of a detailed analysis of the
stellar structure equations. Contrary to the suggestions by
\cite{1993ApJ...415..767I} and \cite{10.1093/mnras/236.3.505} we find
that a simple analytic answer is possible. Our analysis is inspired by
the physical insight offered by \cite{2005slfh.book..149F} but it is
based on a more solid logical and mathematical derivation, offering 
deeper insight into the structure of red giants. Most importantly, our
analysis done in Section \ref{sec:aboveshell} highlights the role
played by convection in the development of (bright) red
giants. Contrary to the conclusion of \cite{2005slfh.book..149F}, that
convection only acts to make stars smaller (see item $(e)$ in Appendix
10.A.2 of \citealt{2005slfh.book..149F}), we find that the development
of convection is a key ingredient. It is the development of convection
what sets the upper boundary to the value of $\nabla_+$, leading to the
existence of the $T_s(M_c, R_s)$ relation. This fact explains why
earlier models in which convection was neglected did not become
luminous red giants---see for example Fig. 3 in
\cite{1952ApJ...116..463S} but also Fig. 2 of
\cite{2009PASA...26..203S}. The fact that \cite{2005slfh.book..149F}
is able to derive a core mass-luminosity relation (his equation 10.7)
with no mentioning of the role of convection highlights the problem
with the mathematical derivations in that work, and in particular with
the analysis done in their Appendix 10.A.2.

Our findings can be summarized as follows.  The existence of an
envelope of nonnegligible mass forces $\nabla$ to have a very narrow
range of possible values ($0.23\lesssim \nabla \leq 0.4$, for typical
opacity laws), which leads to a tight dependence of the shell
temperature on the mass and radius of the core $T_s\propto M_c/R_s$
(eq. \ref{eq:T+_MR}). Additionally, the switching-on of the burning
shell and the fact that the heat is both generated and transported at
the burning shell leads to a tight relationship between $\rho_s$,
$T_s$, $M_c$, and $R_s$ (eq. \ref{eq:the-equation}). Together,
eqs. \ref{eq:T+_MR} and \ref{eq:the-equation} imply that $\rho_s(M_c,
R_s)$ (and consequently $P_s(M_c, R_s)$). In a full stellar
structure, these constraints are then complemented by mass radius
relations for the core $M_c(R_s)$. The addition of this restriction
leads to unique values for $T_s$, $\rho_s$ and $P_s$ as a function of
either $R_s$ or $M_c$. This forces a very different evolution for
$\rho_s$ and $P_s$ as a function of either $R_s$ or $M_c$ in comparison
to their core counterparts $\bar{P_c}$ and $\bar{\rho_c}$, making
homologous contraction impossible.

Although the $M_c(R_s)$ relation of the cores is different in low-mass stars
and in intermediate-mass and massive (IM\&M) stars, the consequences are
the same. In low-mass stars the combination of a degenerate massive
inner region and a massless hot mantle below the burning shell leads
to an almost constant value of $R_s$ while the mass of the core
increases as nuclear burning consumes the H envelope. On the contrary,
on IM\&M stars the Kelvin-Helmholtz contraction of the classical ideal
gas in the core leads to the decrease in the radius of the core but at
almost constant mass. In both low-mass and IM\&M stars, however,
evolution leads to an increase in the mean pressure $\bar{P_c}$ and
density $\bar{\rho_c}$ of the core. Most importantly, evolution in
both low-mass and IM\&M stars leads to an increase in the pressure
and density contrasts between the shell and the core ($\bar{P_c}/P_s$
and $\bar{\rho_c}/\rho_s$). This increase in $\bar{\rho_c}/\rho_s$
leads to a huge expansion of a mantle on top of the burning shell, where
both the pressure and density drop by orders of magnitude. The massive
envelope on top of this mantle is then forced to very low densities
and low gravitational potentials, leading to very large
stellar radii. The storage of a massive envelope at low densities
leads to the formation of a red giant.

Besides finding a toy-model description for the formation of red
giants some additional findings of the present work are worth
mentioning. First, the $T_s (M_c,R_s)$, $\rho_s(M_c,R_s)$, and
$P_s(M_c,R_s)$ relations derived from the analysis of the burning
shell (and the lower envelope) are, to the best of our knowledge, the
first mathematical demonstration of this key hypothesis behind
shell homology relations \citep{2012sse..book.....K}. Second, while it
is known since the work of \cite{1947PThPh...2..127H} that stellar
models become giants even when convection is suppressed, our simple
model shows that convection plays a key role in the formation of the
actual luminous red giants that we observe. Next, it is clear from the
current presentation why a molecular weight gradient helps the
development of a giant structure, although it is neither a sufficient
nor a necessary condition \citep{2009PASA...26..203S}. The development
of a weight gradient between the core and the shell helps increase
the density contrast $\bar{\rho_c}/\rho_s$ between the core and the
envelope, but it is not the only way to attain a large density
contrast. In addition, our toy model shows that it is not completely
correct to estimate changes in the luminosity of the shell due to
changes in the molecular weight without including the feedback of this
change on the radius of the shell \citep[e.g.][]{2012sse..book.....K}. 
Finally, our simple toy model of a low-mass red giant
offers the first simple explanation of why all low-mass stars develop
a He-core flash at basically the same core mass, and why this mass is
of the order of half a solar mass. In a star with an inert degenerate
core the temperature of the shell solely depends on the mass of the
core. In particular, this temperature reaches He-burning luminosities
at $M_c\simeq 0.75 M_\odot$, independently of the initial mass or
metallicity. This sets a clear maximum value for the mass of the
He core at the He flash. The helium flash develops well before this
point because, for the high temperatures of the burning shell, the
nuclear timescales became comparable to the timescale for
gravitational contraction. As consequence, gravitational contraction
provides an additional heating source, leading to an advance of the
He flash. Interestingly, as this advance is a consequence of the rate
of growth of the degenerate core, and that rate ($dM_c/dt\propto L_s$)
solely depends on the mass of the core, the helium flash is advanced
by the same amount in all low-mass stars.

From a pedagogical perspective, we believe that having a simple
physical model to explain how stars become luminous red giants will
improve our teaching of stellar evolution, and also our interpretation
of results from detailed numerical models. In future works we will
apply the model presented here to the explanation of other properties
of stellar models like the red giant bump \citep{2015MNRAS.453..666C,
    2020MNRAS.492.5940H} or the critical mass at which stars depart
  from the red giant branch \citep{2008ApJ...674L..49S}.

In closing let us mention that, given the past history of this topic,
we do not expect this paper to end the discussion about the
subject. We hope, however, that the ideas presented in this paper might
offer a different perspective for future discussions.

\acknowledgments M3B warmly thanks J{\o}rgen Christensen-Dalsgaard,
Achim Wei\ss, Tiara Battich, Santi Cassisi, Leandro G. Althaus, Alfred
Gautschy, and the two anonymous referees for their comments and
criticism on different versions of the manuscript that helped to
improve and correct the paper. J{\o}rgen Christensen-Dalsgaard and
Achim Wei\ss\ are particularly acknowledged, the former for his very
detailed feedback and encouragement, and the latter for his patience
with the author during the many discussions about this topic in the
last 15 years. Tiara Battich is also acknowledged for her continuous
support during the realization of this work. M3B is partially
supported by PIP 2971 from CONICET and PICT 2020-03316 from Agencia
I+D+i. The author also thanks the Max Planck Institute for Astrophysics
without whose library this research would have been impossible. The
author also thanks I. Newton for the idea of what to do during a
pandemic lockdown. Finally, M3B is especially grateful to the editor
(Dr. Steven Kawaler) for his patience and positive approach to the
peer review process.  This work is dedicated to Mabel A. Bertolami who
taught the author the values of hard work and perseverance, which were
key for the successful realization of this work.

%

\vspace{5mm}


\software{{\tt LPCODE:} \cite{2016A&A...588A..25M}}



\appendix

\section{The inner and upper mantle in detailed numerical models}
\label{app:numerical_mantles}

\begin{figure}
\centering
\includegraphics[width=\columnwidth]{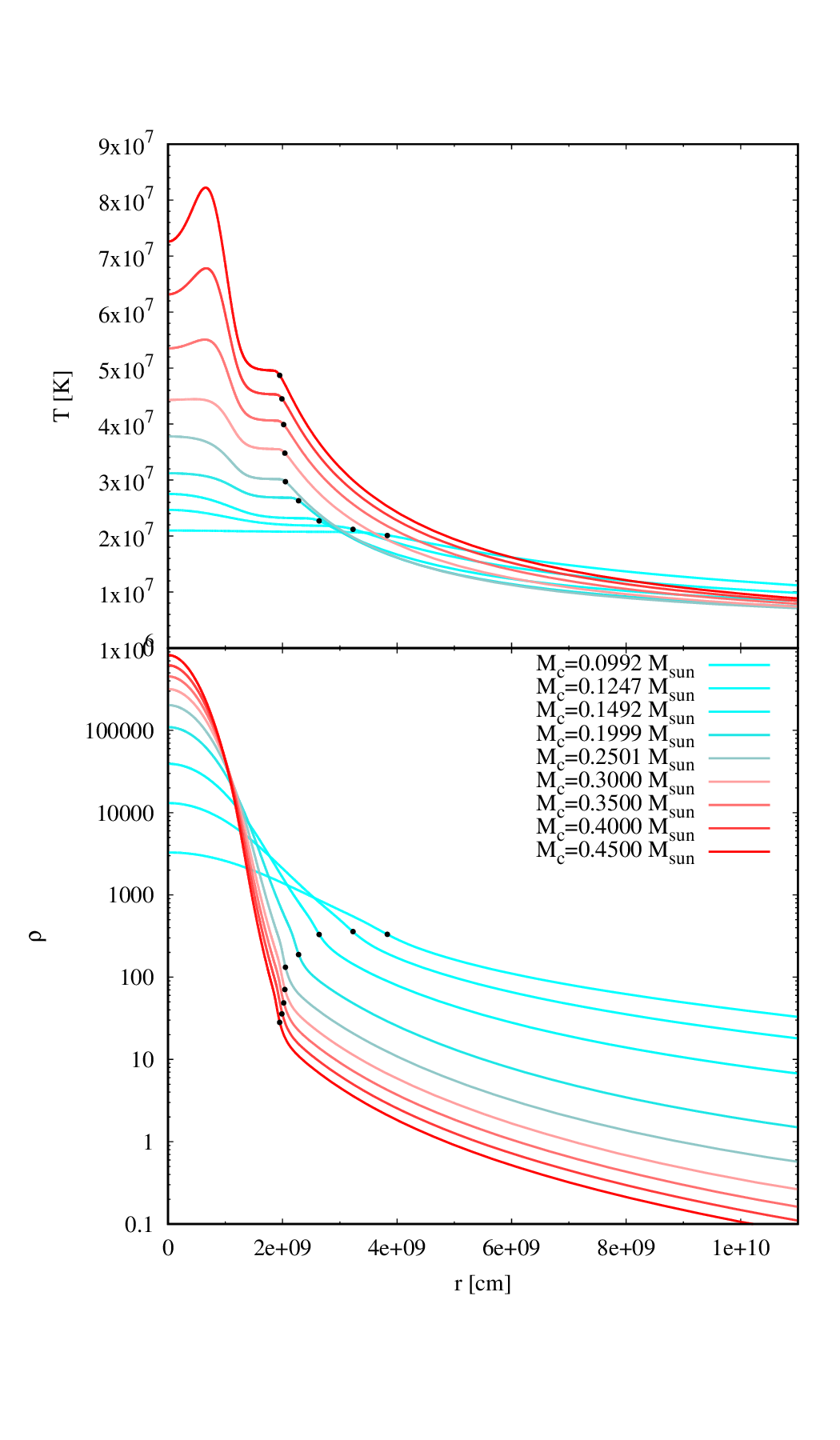}
\caption{Density and temperature profiles of a $1 M_\odot$ model
  ($Z=0.02$) for different core masses. Black dots indicate the
  location of the peak of the H-burning shell of each model.}
\label{fig:modelos}
\end{figure}

In this section we show the main structural properties of the core,
shell, and upper mantle in a detailed numerical simulation of a
$1M_\odot$, $Z=0.02$ star on the red giant branch. We do this for the
sake of completeness and also to show the reasons behind the
identification of the main parts of a low-mass red giant
(Fig. \ref{fig:Structure}).  Fig. \ref{fig:modelos} shows the
temperature and density structure of a detailed $1M_\odot$ model as
it evolves on the red giant branch.

\begin{figure}
\centering
\includegraphics[width=\columnwidth]{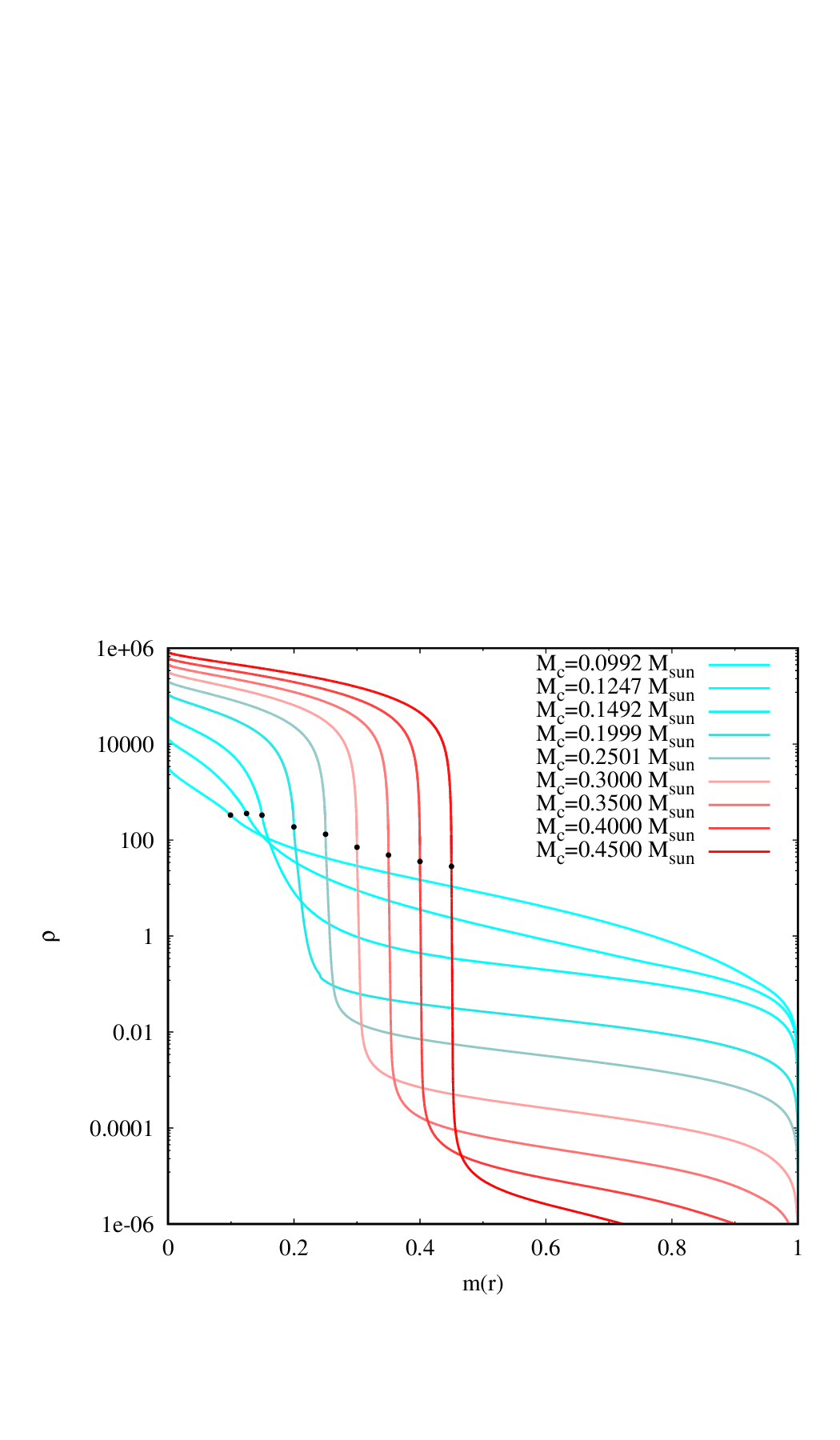}
\caption{Density profiles of a $1 M_\odot$ model ($Z=0.02$) as a
  function of the lagrangian coordinate $m(r)$. Black dots indicate
  the location of the peak of the H-burning shell of each model.}
\label{fig:lagrangian}
\end{figure}

The existence of the inner isothermal mantle is apparent in all models
right below the burning shell (upper panel of
Fig. \ref{fig:modelos}). As expected from the discussion in our toy
model, in this region the density drops exponentially with increasing radius,
from the typical values in the degenerate core down to the density of
the burning shell (lower panel of Fig. \ref{fig:modelos}). Note that,
although extended in radius, the upper mantle region shown in Fig.
\ref{fig:modelos} harbors a negligible amount of mass, due to its low
densities in comparison to the core. This is clearly appreciated in
Fig. \ref{fig:lagrangian} where the properties of the models are shown
as a function of the lagrangian coordinate $m(r)=\int_0^r 4 \pi r'^2
\rho dr'$. In this figure it is clear that, as soon as the star develops a
relatively dense core with a significant mass ($M_c\gtrsim 0.2
M_\odot$), the density drops orders of magnitude in a region of negligible
mass above the burning shell (our so-called upper mantle), as a
consequence of eq. \ref{eq:dTdt-transport-hydro}.

A selected sample of the properties of the models at different stages
of the evolution is shown in Table \ref{tab:1Msun}.

\begin{table}[]
  \begin{tabular}{lccccc}
    $M_c$	 &     $R_c$ &	$T_s$&	$\rho_s$&	$R_\star$&	$\log L/L_\odot$ \\
    $[M_\odot]$   &     [$10^9$ cm]  &   [$10^7$ K] &   [g/cm$^3$] & [$R_\odot$] & \\\hline
  0.0317	 &     3.30 &	1.88 &	314`	&	1.286	&	0.256 \\
  0.0489	 &     3.65 &	1.90 & 	318	&	1.367	&	0.295 \\
  0.0788	 &     3.86 &	1.97 &	332	&	1.519	&	0.355 \\
  0.0992	 &     3.83 &	2.01 &	332	&	1.646	&	0.380 \\
  0.1247	 &     3.23 &	2.12 &	358	&	1.916	&	0.378 \\
  0.1492	 &     2.64 &	2.27 &	330	&	2.543	&	0.528 \\
  0.1752	 &     2.39 &	2.44 &	247	&	3.722	&	0.826 \\
  0.1999	 &     2.28 &	2.63 &	188	&	5.574	&	1.133 \\ 
  0.2501	 &     2.05 &	2.97 &	132	&	10.51	&	1.593 \\
  0.3000	 &     2.04 &	3.48 &	70.9	&	26.17	&	2.222 \\
  0.3500	 &     2.02 &	3.99 &	48.7	&	53.87	&	2.698 \\
  0.4000	 &     1.99 &	4.45 &	35.8	&	96.72	&	3.059 \\
  0.4500	 &     1.95 &	4.87 &	28.2	&	156.5	&	3.336 \\
  \end{tabular}
  \caption{Properties of the shell and the star for selected snapshots
    in the evolution of the $1 M_\odot$ ($Z=0.02$) sequence.}
  \label{tab:1Msun}
\end{table}

\begin{figure*}
\centering
\includegraphics[width=\columnwidth]{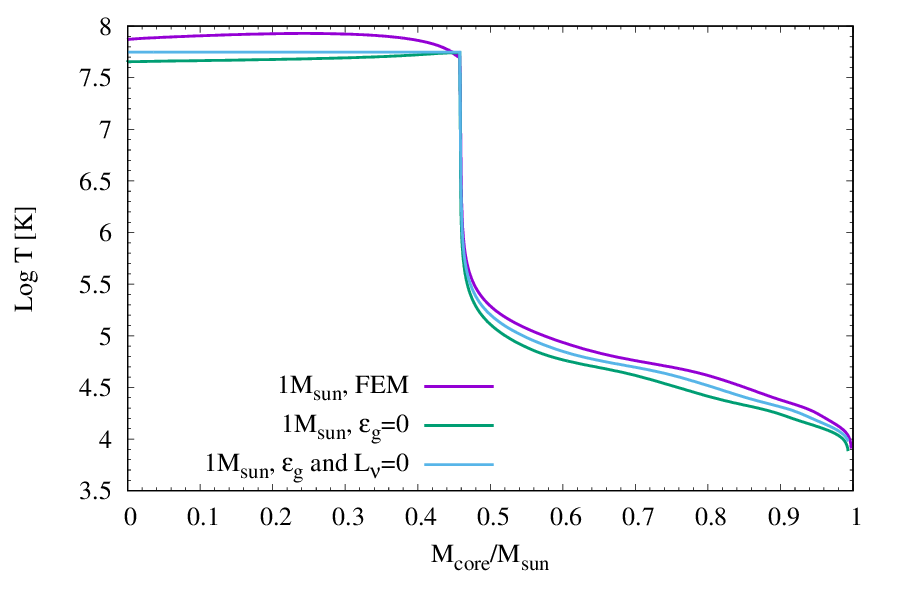}
\caption{Temperature stratification of the $1M_\odot$ full evolutionary
  model (FEM) near the He-flash. Green and cyan lines indicate the
  corresponding temperature stratification when the same chemical
  structure is fixed and the gravothermal energy release is relaxed to
  zero $\epsilon_g=0$ (green), and when in addition to setting
  $\epsilon_g=0$ neutrino energy losses are also set to zero.  }
\label{fig:Relax}
\end{figure*}

A clear distinction between our toy model and the actual solution of
the detailed models is the fact that the cores of red giants are
clearly not isothermal for two different reasons. On the one hand the
material is not completely degenerate ($T=0$) as it is still too hot
for that approximation in the outermost regions of the degenerate
core, and on the other hand the existence of neutrino emission leads
to a decrease in the temperature of the core close to the center. Fig
\ref{fig:Relax} shows a full evolutionary model near the He-core flash
in comparison to the corresponding structure when gravothermal energy
release is relaxed to zero (for the same chemical structure). Clearly,
in thermal equilibrium the degenerate core attains a temperature
almost equal to that of the burning shell, with a slight decrease near
the center of the core, due to neutrino energy losses. If neutrino
losses are also removed, then the core becomes completely isothermal
(Fig. \ref{fig:Relax}). Fig. \ref{fig:Relax} shows that the He flash
happens for a lower core mass than it would were it the case that
nuclear reactions had a much larger timescale than the
Kelvin-Helmholtz timescale of the core.

\section{Relations for a Kramers' bound-free opacity law}
\label{app:opac}

In Section \ref{sec:shell} we derived the equations for the upper
mantle under the assumption that opacity near the burning shell is
dominated by classical electron scattering. While this is true once
$M_c\gtrsim 0.23 M_\odot$ (for a $1M_\odot$ star), below that core
mass (and the consequent shell temperature) other sources of opacity,
mostly bound-free opacities, are important. To explore the impact of
this assumption we rederive the equations used in Sections
\ref{sec:shell} and \ref{sec:toystory} when Kramers' bound-free opacity is adopted.

A very rough approximation to bound-free opacities in stellar
interiors is provided by \cite{2004sipp.book.....H}, which for a typical composition of the envelope ($X=0.7$, $Z=0.02$) is (in cgs)
\begin{equation}
  \kappa_{\rm b-f}=4.3\times 10^{-8} \rho {T_9}^{-7/2}.
  \label{eq:kram}
\end{equation}  
With this new opacity law we have $a=1$ and $b=-4.5$, and
eqs. \ref{eq:OuterPT}, \ref{eq:burningshell} can be derived in exactly
the same way as before.  Using these equations and the opacity law in
eq. \ref{eq:kram} we can now derive $R_s(M_c)$ and $T_9(R_s)$ relation similar
to eqs. \ref{eq:xmT9} and \ref{eq:xT9}. These new relations are
\begin{equation}
  \frac{R_s}{R_{\rm dc}}\simeq \frac{\mu_{\rm env}}{\mu_c}\frac{0.56}{T_9}
  \left(\frac{M_{\rm dc}}{M_\odot}\right)^{4/3},
  \label{eq:xmT9_kram}
\end{equation}
and
\begin{equation}
 \begin{aligned}  
 1.90\times & 10^{-25} \frac{\mu_c^4}{\mu_s^2\mu_{\rm env}^2}\nu \exp(30/\nu)\left(1+1/\nu\right)^{-6}=\\& {T_9}^{-13/6} \left[\frac{R_s}{R_{\rm dc}}\right]^2\left[\frac{M_c}{M_\odot}\right]^{-2/3} \\&
 \exp\left[\frac{17}{(1+2/\nu)}\frac{\mu_c}{\mu_{\rm env}}\left(1-\frac{R_{\rm s}}{R_{\rm dc}}\right)-15.231 {T_9}^{-1/3}\right]
 \end{aligned} 
 \label{eq:RT_kram}
\end{equation}
The resulting $T_9(M_c)$ $R_s(M_c)$ are shown in Fig. \ref{fig:RT}
where it is clear that they share the same key features described by
eqs. \ref{eq:xmT9} and \ref{eq:xT9}, the burning shell (and
consequently the isothermal core below) attains  He-burning temperatures
at $M_c\simeq 0.7 M_\odot$ and the H-burning shell settles at a
constant radius of $R_s\simeq 0.028 R_\odot$, very close to that
predicted by full stellar evolution models.

The main difference introduced by a Kramers' opacity law is to increase the temperature of the shell as compared to
that predicted by Thomson scattering for the more massive
cores. Note, however, that as soon as the temperature of the shell
increases to $T\gtrsim 3\times 10^7$ K, electron scattering becomes
the main opacity source. 

For the sake of completeness let us mention that the solutions shown in  Fig. \ref{fig:RT} can be very well approximated by

\begin{eqnarray}
  T_9&=&0.0037+0.136 \left(\frac{M_c}{M_\odot}\right)^{1.16}\nonumber\\
  \frac{R_s}{R_{\rm dc}}&=&1.884+ 0.357 \,\ln \left(\frac{M_c}{M_\odot}-0.039\right).
 \label{eq:TR_M-krammers} 
\end{eqnarray}

\section{Very deep convection}\label{app:deep_conv}
\begin{figure}
\centering
\includegraphics[width=\columnwidth]{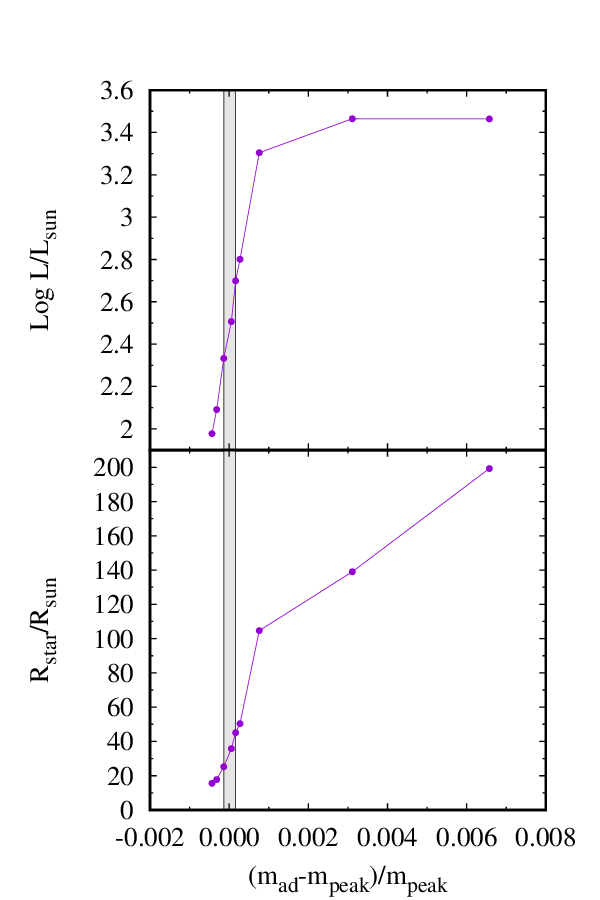}
\caption{Impact of very deep convection on the luminosity and radius
  of the star.}
\label{fig:Rho-M-kram}
\end{figure}

Key equations derived in the paper rely on the radiative nature of the
 burning shell. Here we analyze how expressions presented in the
paper are affected by convection. Let us call $L_s$ and $L_c$ the
luminosities of the shell and core (where $L_s\gtrsim L_c$).
Let us assume now that, for some reason, convection reaches deep into
the burning shell to a point where the luminosity is
$L_\chi=L_s\,\chi$ with $0<\chi<1$. At the convective boundary we have that $\nabla_{\rm ad}=\nabla_{\rm rad}$, and, by equating the temperature gradients we have
\begin{equation}
\left.\frac{dT}{dr}\right|_{\chi} = -0.4\frac{G M_c \mu}{\Re {R_s}^2},
\label{eq:conv1}
\end{equation}  
and from the energy generation and energy transport equations we can write
\begin{equation}
\left(\left.\frac{dT}{dr}\right|_{\chi}\right)^2 \simeq
  \frac{3}{8\pi a c}\frac{{\rho_s}^2}{{T_s}^2} \frac{\kappa_s \epsilon_s}{\nu}
  \left(\frac{L_c}{L_s}+\chi\right) \mathcal{D},
  \label{conv2}
\end{equation}
where $1<\mathcal{D}<2$ if convection has not reached the location of the peak of the burning shell and $1=\mathcal{D}$ if it has. Together eqs. \ref {eq:conv1} and \ref{conv2} give
\begin{equation}
  \left(0.4\frac{G M_c \mu}{\Re {R_s}^2}  \right)^2\frac{1}{\left(\frac{L_c}{L_s}+\chi\right)}
  \approx
  \frac{3}{8\pi a c}\frac{{\rho_s}^2}{{T_s}^2} \frac{\kappa_s \epsilon_s}{\nu}\mathcal{D}.
  \label{conv3}
\end{equation}
Although the equations remain formally unchanged, it is clear that the
deepening of convection inside the burning shell will have a strong
impact. As convection deepens, then $\chi\longrightarrow 0$ and the
left- and right-hand sides of eq.  \ref{conv3} are decoupled. Without
this constraint all the arguments that allowed us to conclude that a
red giant will be formed cannot be stated. In fact, if somehow a
steady burning shell would develop in a completely convective star,
that object would not develop a giant structure, as those stars are
$n=3/2$ polytropes and, as such, cannot develop a density contrast
between the envelope and the core. It is clear that convection inside
the burning shell will conspire with the formation of a giant
structure.  Moreover, numerical experiments show that, as soon as
convective transport penetrates inside the main regions of the burning
shell, the radius of the star will decrease.
Fig. \ref{fig:Rho-M-kram} shows a numerical experiment in which the
bottom of the adiabatic temperature is forced down into the burning
shell itself. Note that these are static structures in which the
gravothermal heat term $\epsilon_g$ is forced to zero throughout the
star. We see that, as long as the adiabatic gradient stays far from the
burning shell, the luminosity of the star remains unchanged because
both the upper mantle and the burning shell can be considered
radiative so the temperature of the burning shell and luminosity are
determined by the equations derived in this paper. The radius of the
star, on the other hand, is reduced as a consequence of the less steep
density gradient imposed by adiabatic convection. However, as soon as
convection reaches down to the burning shell itself luminosity is
strongly affected, and the radius of the star drops suddenly. The more
convection reaches into the burning shell, the smaller the final
radius of the star. We find that, although relatively deep convection
remains the main picture of the paper, as long as convection does not
reach the bottom of the burning shell, stars with deep convective
zones in the sense of this section (i.e. those in which convection
reaches down to the burning shell itself) display much smaller radius.

\section{Envelope Integrations}
\label{app:envelopes}

\begin{figure}
\centering
\includegraphics[width=\columnwidth]{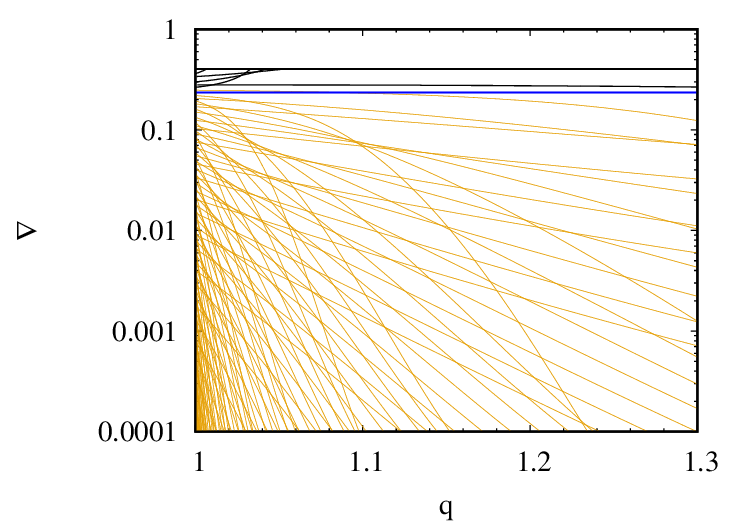}
\includegraphics[width=\columnwidth]{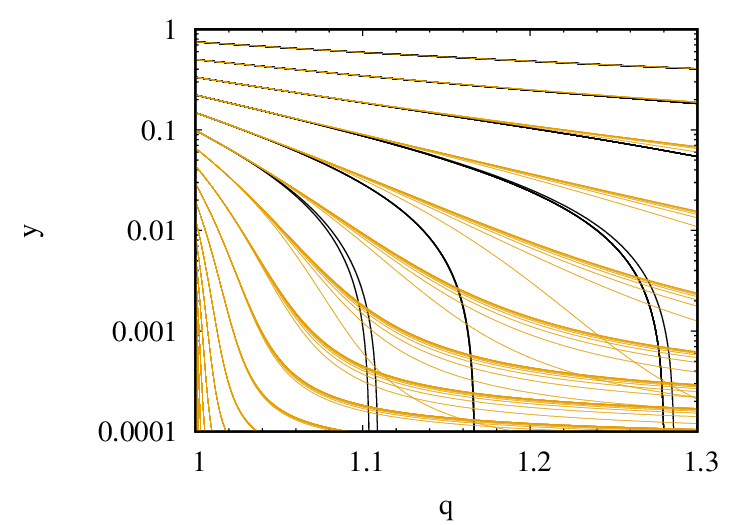}
\includegraphics[width=\columnwidth]{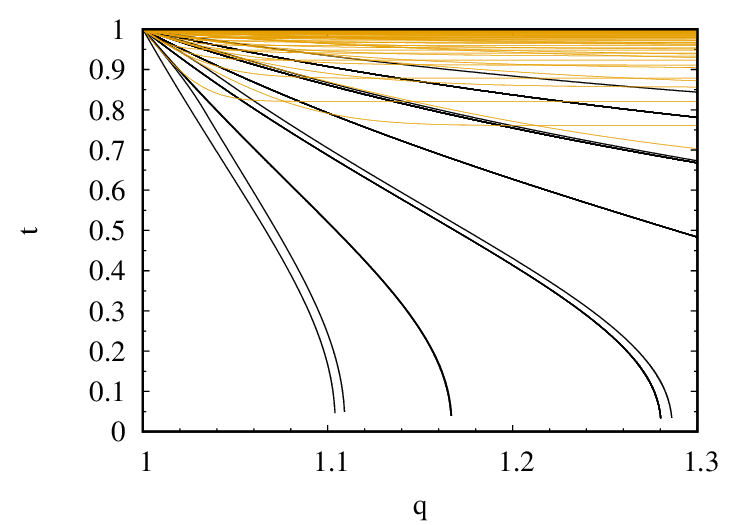}
\caption{Envelope integration for different values of $\mathbb{C}$ and $y_+$, for $t_+=1$ and for a Kramers' opacity.}
\label{fig:envolturas2}
\end{figure}

\begin{figure}
\centering
\includegraphics[width=\columnwidth]{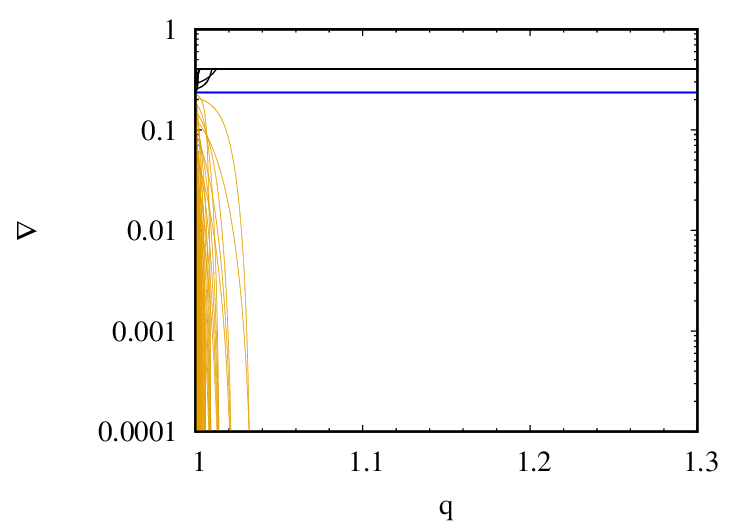}
\includegraphics[width=\columnwidth]{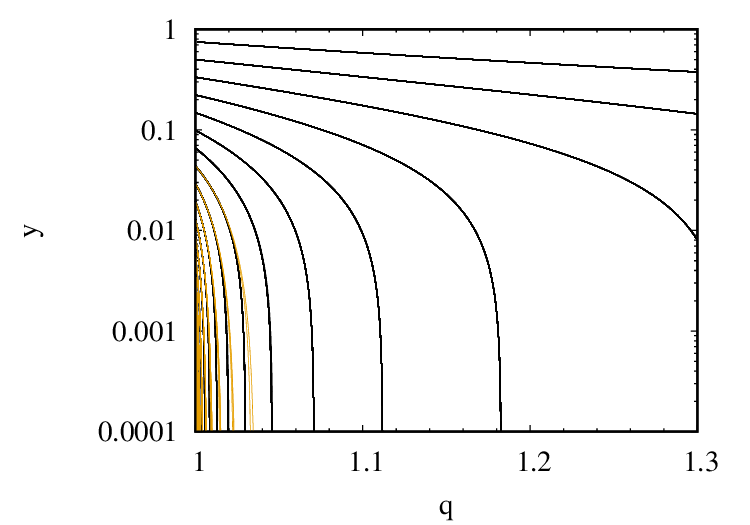}
\includegraphics[width=\columnwidth]{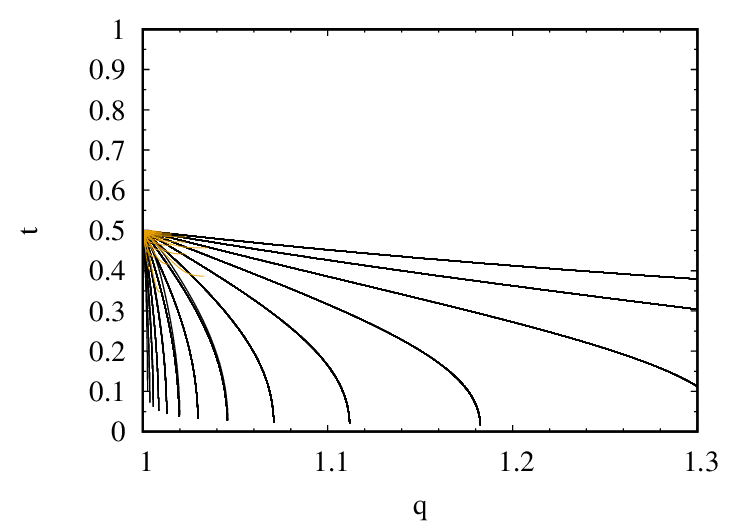}
\caption{Envelope integration for different values of $\mathbb{C}$ and $y_+$, for $t_+=0.5$ and for a Kramers' opacity.}
\label{fig:envolturas3}
\end{figure}

\begin{figure}
\centering
\includegraphics[width=\columnwidth]{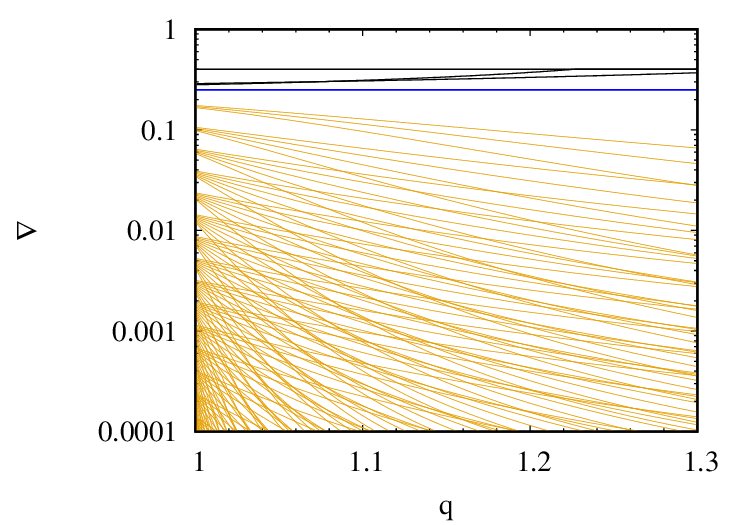}
\includegraphics[width=\columnwidth]{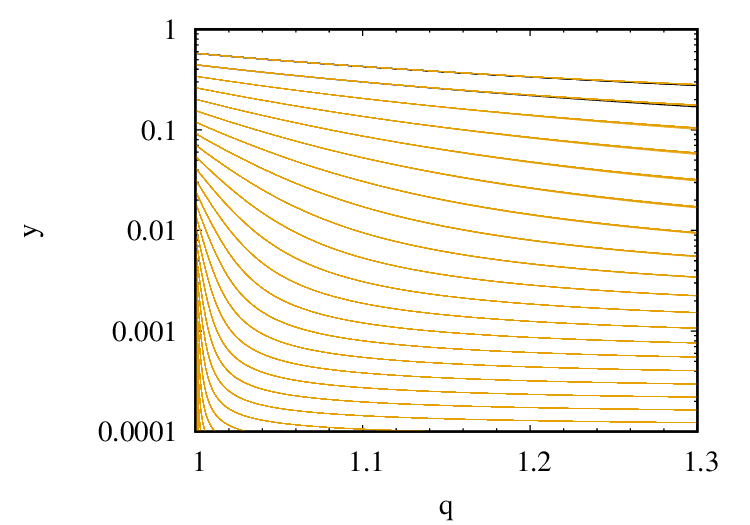}
\includegraphics[width=\columnwidth]{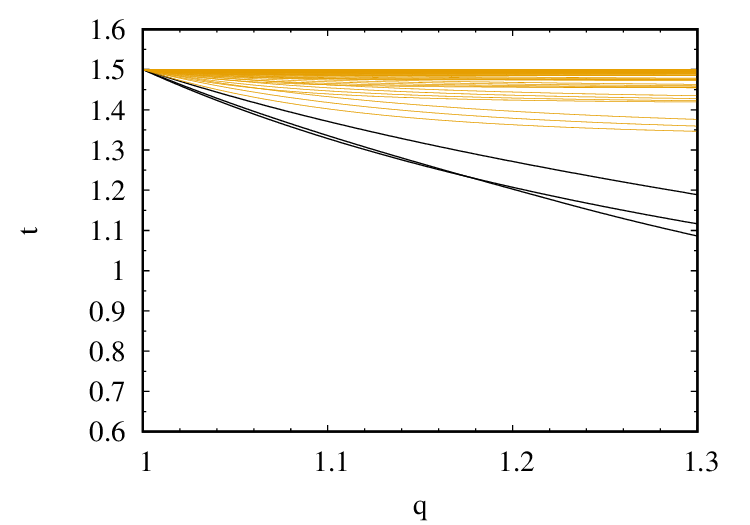}
\caption{Envelope integration for different values of $\mathbb{C}$ and $y_+$, for $t_+=1.5$ and for a Kramers' opacity.}
\label{fig:envolturas4}
\end{figure}

\begin{figure}
\centering
\includegraphics[width=\columnwidth]{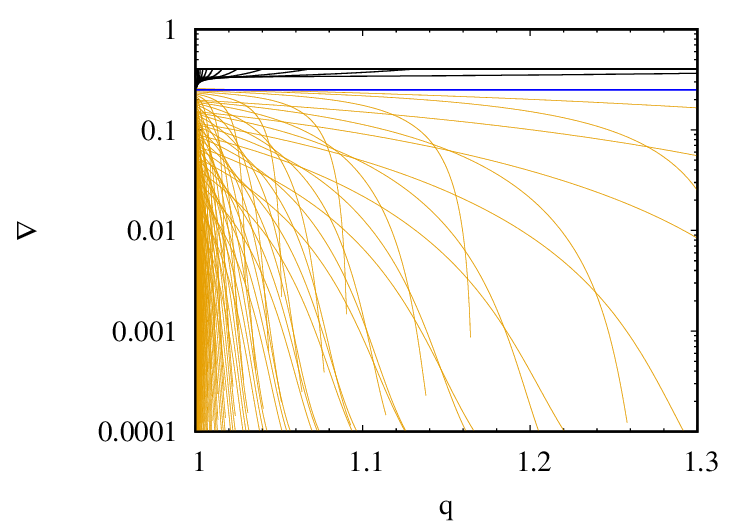}
\includegraphics[width=\columnwidth]{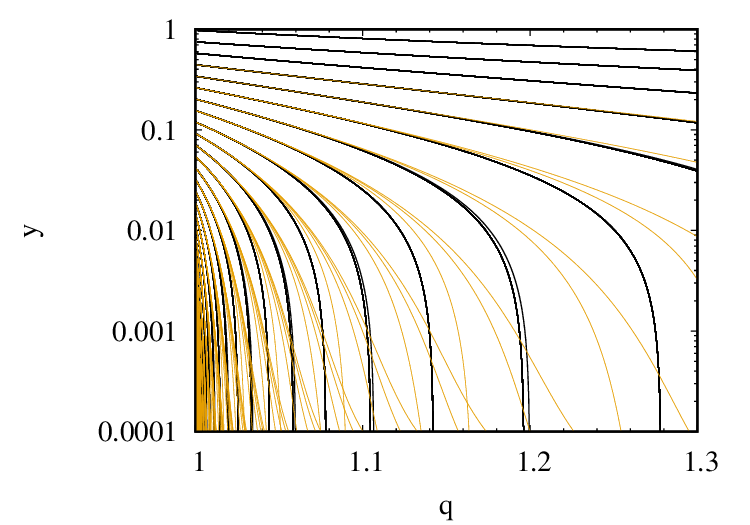}
\includegraphics[width=\columnwidth]{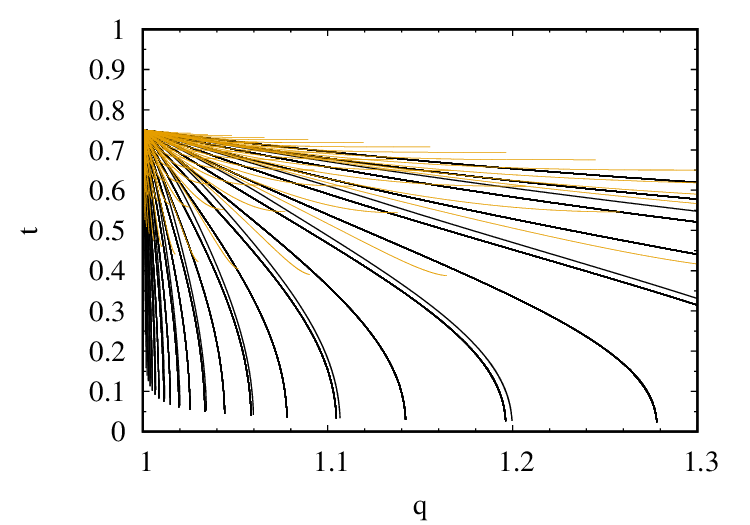}
\caption{Envelope integration for different values of $\mathbb{C}$ and $y_+$, for $t_+=0.75$ and for a Thomson scattering-like opacity.}
\label{fig:envolturas5}
\end{figure}

Figs. \ref{fig:envolturas2}, \ref{fig:envolturas3},  \ref{fig:envolturas4}, and  \ref{fig:envolturas5} show envelope integrations similar to those displayed by Fig. \ref{fig:envolturas} but for different conditions at the burning shell. 

\section{Early evolution of the burning shell}
\label{app:early}

As discussed in Section \ref{subsec:upper}, when $\rho_+\ll \rho_c$
and $P_+\ll \bar{P_c}$ we can strictly demonstrate that $\Theta\simeq 1$,
$\Pi\simeq 1$, $\zeta\simeq 1$, $T_0\ll T_+$ and $R_s\ll R_0$ and all
previous expressions are very simple. We show below that these
approximations are still acceptable even on the very early stages of
shell burning. From the hydrostatic equilibrium equation and using the
definitions in eq. \ref{eq:characteristic_PrhoT} we can see that the
drop in pressure, temperature, and density immediately above the
burning shell follows
\begin{equation}
 \begin{aligned}  
  &\left. \frac{dP}{P}\right|_+ \simeq -6 \left(\frac{T_c}{T_+}\right) \frac{R_s\, dr}{r^2},\\ 
&\left. \frac{dT}{T}\right|_+ \simeq -6 \nabla_+ \left(\frac{T_c}{T_+}\right)\frac{R_s\, dr}{r^2},\\ 
&\left. \frac{d\rho}{\rho}\right|_+ \simeq -6 \left(1-\nabla_+\right) \left(\frac{T_c}{T_+}\right)\frac{R_s\, dr}{r^2}.
\end{aligned} 
\label{eq:drop_PTrho}
\end{equation}  
From eqs. \ref{eq:estimacion_P_Rho} we see that once the shell becomes active $F\simeq 2$ (eq. \ref{eq:inside3}) in the early stages of H-shell burning ($\nu\simeq 23$ at low temperatures of the CNO cycle), we have for $\nabla_+\simeq 0.25$ very roughly that $\rho_+ \sim \rho_-/2$ and $P_+ \sim P_-/2$. Using the very conservative approximation that at the lower boundary of the burning shell density and pressure are half their values at the core, we then have $\rho_+ \sim \rho_c/4$ and $P_+ \sim \bar{P_c} /4$ (in fact, in low-mass giants it will be much smaller than this due to the presence of a near isothermal inner mantle; see Section \ref{sec:innermantle}). From eqs. \ref{eq:drop_PTrho} and as long as $r\simeq R_s$ we can make the very rough approximation that $d\rho/\rho|_+\simeq -18/4 dr/r$ immediately above the burning shell. From the previous
estimation we get that $\rho(r)\simeq \rho_+
(r/R_s)^{-18/4}$. Integrating the upper mantle
\begin{equation}
  \mathcal{Q} M_c = \int_{R_s}^{R_{\rm 0}} \rho(r) 4\pi r^2 dr
  \label{eq:density0}
\end{equation}  
where $\mathcal{Q}$ is the fraction of the mass of the core we
consider acceptable for $m(r)\simeq M_c$. Integrating out to
$\mathcal{Q}=0.2$ so that the approximation $m\simeq M_c$ is accurate
to $\sim 20$\%, we get that $R_{\rm 0}\simeq 1.4\, R_s$, or the width
of the upper mantle to be $\Delta R/R_s\simeq 0.4$. From this
estimation we see that, already at the very early stages of H-shell
burning the factor $(1-R_s/R_{\rm 0})$ in eq. \ref{eq:T+_MR} is a
factor three lower than unity but not negligible. From this value of
$\Delta R/R_s$ and the estimations of $dP/P$ and $dT/T$ at the upper
boundary we get $T_{\rm 0}\sim T_+ \exp(-6\nabla_+) (R_0/R_s)\sim 0.31
T_+$ and $P_{\rm 0}\sim \exp(-6) (R_0/R_s)\sim 0.0035 P_+$. Then the
factors $\Theta$ and $\Pi$ in eq. \ref{eq:OuterPT} are both close to
one even at this early stage and considering these conservative
assumptions regarding the link between the values at the lower
boundary ($\rho_-$, $P_-$) and their mean values at the core.

The importance of the estimations from the previous paragraphs cannot
be downplayed as they tell us that already from the very early stages
of H-shell burning the factors $\Pi$, $\Theta$, and $(1-R_{\rm
  0}/R_s)$ in eqs. \ref{eq:OuterPT} and \ref{eq:T+_MR} are of order
unity, and consequently, these equations are actual restrictions for
the values of $\bar{P_c}$, $T_c$ and $R_c$. These restrictions play a key
role in the increase in the density and pressure contrast
($\bar{\rho_c}/\rho_s$ and $\bar{P_c}/P_s$) between the shell and the
core, so it is important to show that such relations are valid (even
if only approximately) before the values of $\bar{\rho_c}/\rho_s$ and
$\bar{P_c}/P_s$ become huge. Once $\bar{\rho_c}/\rho_s$ and
$\bar{P_c}/P_s$ become large, the approximation $\Theta=1$, $\Pi=1$
becomes very precise.  It is then reasonable to assume for most practical estimations that $\Theta\simeq 1$, $\Pi\simeq
1$,$\zeta\simeq 1$ and $R_s\ll R_0$.

\bibliography{biblio}{}
\bibliographystyle{aasjournal}



\end{document}